\documentclass[12pt,fleqn]{article}

\usepackage{graphicx}
\usepackage[labelformat=simple]{subcaption}
\usepackage{dsfont}
\usepackage{relsize}
\usepackage{algorithm}
\usepackage{algpseudocode}
\usepackage{setspace}
\usepackage{amsmath}
\usepackage{amsthm}
\usepackage{amsfonts}
\usepackage{multirow}
\usepackage{array}
\usepackage{booktabs}
\usepackage{amssymb}
\usepackage{color}
\usepackage{authblk}
\usepackage{bbm}
\usepackage{url}
\usepackage{float}
\usepackage{afterpage}
\usepackage[titletoc,title]{appendix}
\usepackage[dotinlabels]{titletoc}
\algrenewcommand\alglinenumber[1]{\footnotesize #1}

\newcommand{\bv}{\mbox{\boldmath$\beta$}}
\newcommand{\tv}{\mbox{\boldmath$\theta$}}
\newcommand{\sigv}{\mbox{\boldmath$\sigma$}}
\newcommand{\Sv}{\mbox{\boldmath$\Sigma$}}

\newcommand{\bbar}{\mbox{$\bar \beta$}}
\newcommand{\hb}{\mbox{$\hat \beta$}}
\newcommand{\hbv}{\mbox{\boldmath$\hat \beta$}}
\newcommand{\hbo}{\mbox{$\hat \beta_{\rm orig}$}}
\newcommand{\hbr}{\mbox{$\hat \beta_{\rm rep}$}}

\newcommand{\br}{\mbox{$\beta_{\rm rep}$}}

\newcommand{\Xv}{\mbox{\boldmath$X$}}
\newcommand{\sgn}{\mbox{sgn}}

\newcommand{\Mr}{{\rm M_R}}

\newtheorem{proposition}{Proposition}

\title{{\bf Statistical Assessment of Replicability via Bayesian Model Criticism}}
\author{Yi Zhao\thanks{zhayi@umich.edu}\,  and Xiaoquan Wen\thanks{xwen@umich.edu}}
\affil{Department of Biostatistics, University of Michigan, Ann Arbor, MI, USA}
\date{}
\begin{document}
\maketitle

\begin{abstract}
Assessment of replicability is critical to ensure the quality and rigor of scientific research. In this paper, we discuss inference and modeling principles for replicability assessment. Targeting distinct application scenarios, we propose two types of Bayesian model criticism approaches to identify potentially irreproducible results in scientific experiments. 
They are motivated by established Bayesian prior and posterior predictive model-checking procedures and generalize many existing replicability assessment methods. Finally, we discuss the statistical properties of the proposed replicability assessment approaches and illustrate their usages by simulations and examples of real data analysis, including the data from the Reproducibility Project: Psychology and a systematic review of impacts of pre-existing cardiovascular disease on COVID-19 outcomes. 
\end{abstract}

\newpage

\section{Introduction}

Reproducibility is a hallmark of scientific research.
It is a necessary characteristic to ensure correctness and rigor for scientific discoveries  \cite{Claerbout1992, Peng2011, Crook2013, Heller2014, Goodman2016, Nichols2017, Haibe2020}.
Despite the rising awareness across all scientific disciplines, there is a general lack of theoretical foundation and methodological work focusing on the statistical assessment of reproducibility/replicability in scientific practice.  
Consequently, inconsistent specifications of {\em replication successes} and misuses of subjective measures for assessing reproducibility are common in scientific practice \cite{Schmidt2009, Plesser2018, Mikowski2018}.
For example, in reporting the findings from Open Science Collaboration's Reproducibility Project: Psychology (RP:P) \cite{Open2015}, media reports focused on one particular metric that more than half of the replications fail to reproduce statistically significant results in the expected directions and made a sensational statement that ``more than half of the scientific results are not reproducible''.
However, using repeated statistical significance to define replication success in such contexts is considered deeply flawed \cite{Goodman1992, Gilbert2016, Patil2016, Hung2020, Gibson2020}.
Because variability is at the core of reproducibility \cite{Karp2018, Kochunov2019, Zhao2020}, statistics is uniquely positioned for providing sound and rigorous solutions.
In a recent review, He and Lin list ``Statistical Methods for Reproducibility and Replicability'' as one of the ten immediate challenges and opportunities in statistics and data science \cite{He2020}.

In this paper, we focus on a particular mode of reproducibility, known as {\em replicability} or {\em results reproducibility}, according to the lexicon of reproducibility by Goodman {\it et al.} \cite{Goodman2016}.
Roughly speaking, it refers to the consistency of the results from analyzing {\em different} data collected to address the same underlying scientific questions. 
For simplicity, we use the terminology {\em replicability} and {\em reproducibility} interchangeably throughout this paper.

There is a large volume of existing methodological works from relevant fields (e.g., systematic review and meta-analysis) related to replicability assessment \cite{Higgins2002, Magosi2017, Jaljuli2019}.
Some are motivated by investigating effects of irreproducibility due to specific factors, e.g., publication bias/winner's curse \cite{Macaskill2001, Zollner2007,Peters2010, Lin2018,Palmer2017, Hung2020} and batch effects \cite{Johnson2007, Leek2010, Goh2017}, in specific settings.
They utilize different inference strategies, e.g., model comparison \cite{Li2011, Zhao2020}, predictive interval checking \cite{Patil2016, Pawel2020}, and deals with distinct application scenarios. 
This diverse body of work lays a solid foundation for us to summarize important statistical principles in replication assessment. 
In this work, we take a Bayesian model criticism strategy for replicability assessment. 
The rich literature \cite{Box1980, Rubin1984, Meng1994, Gelman1996, Gelman2005, Gelman2013} and the well-developed computational infrastructure on Bayesian model criticism make it feasible and attractive for our applications of interest.

In the rest of the paper, we first discuss statistical considerations for replicability assessment and argue that Bayesian model criticism is appropriate and effective for this purpose. 
We then propose two different model criticism approaches, targeting different application scenarios. Finally,  we illustrate their usages through simulation studies and examples of real data analysis. We conclude the paper by discussing important technical issues in the proposed approaches and their connections to alternative statistical strategies.  

The software implementation of the proposed statistical methods (R package, \texttt{PRP}) and the relevant code to reproduce all the simulation and data analysis results can be downloaded from the GitHub repository: \url{https://github.com/ArtemisZhao/PRP}.

\section{Statistical Considerations for Assessing Replicability}

This section first lays out general modeling and inference principles in defining replication successes and assessing replicability.
Based on these guiding principles, we will further consider different application scenarios and discuss potential statistical solutions.

\subsection{Modeling and Inference Principles}

Consider a typical setup of replicability assessment where different experiments are conducted to address the same scientific questions.
Two distinct sources contribute to the variability of analysis results from different experimental datasets.
First, random noise is intrinsic to each dataset (and generally considered independent across datasets). Second, natural variations of underlying true effects of interest are expected across different experiments.
The observed variations among analysis results reflect their combined effects.
In assessing replicability, we argue that evaluating variability from the second source should be the sole focus.
This principle is consistent with the common practice in quantifying and controlling heterogeneity in meta-analysis and systematic review literature \cite{Higgins2002, Peters2010, Lin2018}.

Even though experiment-specific random noise is considered a nuisance in replicability assessment, the determination of the true effects is inevitably confounded by its existence.
The following question best characterizes the role of random noise: should an extremely noisy replication experiment be considered as evidence against the replicability?
The logical answer seems to be no.
As accurate estimation of true underlying effects becomes infeasible in the presence of a high level of random experimental errors, the replication data are essentially {\em non-informative}, which is fundamentally distinct from evidence against reproducibility.
This line of reasoning establishes an asymptotic behavior for desired replicability assessment approaches, which we refer to as the non-informativeness principle of noisy replications henceforth.
An important corollary from the principle is the requirement of explicit uncertainty specification of observed effects as integral input information for replicability assessment.
In comparison, characterizing signal-to-noise ratios as a compound quantity (e.g., $p$-values derived from pivotal $z$-statistics) may be insufficient and violate the very principle (as the noise level can not be separately recovered).

Establishing an extent of heterogeneity for reproducible effects is another critical aspect for defining replication success.
It is overly unrealistic and unnecessarily restrictive to assume reproducible results must have {\em identical} underlying effects \cite{Higgins2002, Lin2018}.
Ultimately, defining acceptable heterogeneity in successful replications should be context-specific.
Nevertheless, we argue that it is plausible to specify an {\it a priori} minimum requirement of the maximum tolerable heterogeneity generally, which is analogous to applying the minimax principle in decision theory.
Recently, we have proposed such a standard, known as the directional consistency (DC) criterion, which emphasizes that reproducible effects are expected to show concordant (positive or negative) signs across replication experiments with a high probability \cite{Zhao2020}.
Similar ideas have been successfully adopted in general hypothesis testing, meta-analysis, system reviews, and the statistical analysis of qualitative interactions  \cite{Gelman2000, Owen2009, Wen2014, Stephens2016, Jaljuli2019}.
It is worth noting that the implementation of DC the criterion in \cite{Zhao2020}  is not {\em solely} focusing on the variability on a particular direction (i.e., the sign-flipping direction).
It utilizes a variance parameter to define a range of acceptable heterogeneity centering around a true latent effect.
This paper illustrates our proposed statistical methods by applying the DC criterion but note that other suitable definitions for acceptable replication variations are also applicable.

\subsection{Application Scenarios}

In practice, replicability assessment is applied in various distinct scenarios, representing some unique characteristics and requiring context-specific statistical treatments.

In one of the common application scenarios, a designated replication experiment is conducted to validate the result from an original study.
The original and replication labels are typically assigned following the chronicle order of data generation.
The RP:P and Experimental economics replication project \cite{EconRP} studies are representative examples of such kind.
The setup can generally be extended to two groups of experiments (i.e., the original group and the replication group). The scientific question of interest is whether the results derived from the replication group confirm the finding from the original group.
For such a design, the replication assessment should be specific to the assigned labels, i.e., we may expect qualitatively different results if the original and replication labels are switched in the analysis.
We will henceforth refer to this scenario as the two-group scenario.

Another application scenario of replicability assessment often arises from systematic review and meta-analysis, where a single group of multiple experiments is gathered.
The relevant scientific question here is to assess the overall concordant of all experiments and identify potential outlying results.
Notably, the chronicle order of data generation usually plays no role in the data analysis, and switching chronicle labels of the participating experiments should not yield quantitatively different assessment results.
We will refer to this particular scenario as the (chronically) exchangeable group scenario henceforth.

From a statistical perspective, the two distinct scenarios above suggest different statistical treatments in replicability assessment: the analysis in the two-group scenario calls for explicit conditioning of the original experiment, whereas the chronically exchangeable structure among experiments should be preserved in the second scenario.

There are other distinguishing factors in applications of replicability assessment. For example, most experiments in physical and social sciences contain a single analysis unit. However, in biological science, high-throughput experiments, where tens of thousands of biological units (e.g., genes) are simultaneously measured, have become increasingly common.
Thus, the unit- and study-level assessments are no longer synonymous and require specialized statistical treatments \cite{Li2011, Zhao2020}.
This paper will focus on the former case, and we further address their connections and distinctions in the Discussion section.

\subsection{Model Criticism Strategy}

In this paper, we propose to assess replicability by applying a model criticism strategy. Specifically, we formulate a parametric model representing the expected characteristics from reproducible results. Subsequently, we fit such a model with observed data and evaluate the goodness-of-fit.
Specifically, poor-fitting prompts to reject the notation of the observed data are likely reproducible.
Because our employed parametric models are (Bayesian) hierarchical, we apply the principled framework of Bayesian model checking approaches for replicability assessment. Particularly, we utilize both prior checking \cite{Box1980} and posterior checking \cite{Rubin1984, Gelman1996} techniques to deal with different application scenarios.

In our proposed methods, $p$-value is one of the primary statistical instruments to summarize and quantify the results from model criticism procedures.
Our usage of $p$-values in this context closely relates to the Fisherian significance testing and directly follows the applications in Bayesian model checking \cite{Box1980, Rubin1984}.
Specifically, $p$-values quantify the discrepancy between the assumed reproducible model and the observed data, as in the original Fisherian design.
We also adopt Fisher's disjunction to interpret a small $p$-value, which indicates ``either an exceptionally rare chance has occurred or the theory is not true'' \cite{Fisher1956}.
Here, the specific ``theory" refers to the built-in reproducible assumptions.
Additionally, our use and subsequent interpretations of $p$-values in the context of replicability assessment strictly follow the ASA's statement on $p$-values \cite{ASA2016}.

Critically, our use of model criticism approaches and $p$-values should not be confused with the binary decision procedures characterized by the Neyman-Pearson hypothesis testing \cite{Hubbard2003} and the hybrid null hypothesis significance testing (NHST).
Our goal, in this paper, is {\em not} to classify the observed data into reproducible and irreproducible categories. Hence, the $\alpha$ and $\beta$ levels for type I and type II errors are irrelevant in our proposed statistical framework.
On the other hand, we need to elucidate the statistical properties of the proposed procedures in dealing with truly replicable and irreproducible results. Specifically, the calibration of the proposed $p$-values under the reproducible model and the sensitivity of the procedures under various irreproducible scenarios \cite{Hubbard2003}.
We will use dedicated terminologies in such cases to avoid confusion.
The distinction between Fisherian significance testing and Neyman-Pearson style NHST has profound implications in our treatments of some important statistical issues, e.g.,  the simultaneous examinations of multiple replication experiments. The relevant points will be fully illustrated in the subsequent sections.

Notwithstanding the importance of $p$-values in quantifying evidence, our model criticism strategy for replication assessment also utilizes other suitable statistical instruments and visualization tools.
We emphasize that replication assessment via model criticism is a comprehensive data analysis process and should not be simplified into a $p$-value-generating procedure.

\section{Two Model Criticism Approaches}

\subsection{Model and Notation}

Consider $m$ experiments to address the same scientific question. The complete collection of the experiment data are denoted by $(\Xv_1,...,\Xv_m)$.
Let $\hb_i$ and $\sigma_i$ denote the measurement and its corresponding standard error of the underlying effect of interest from the $i$-th experiment, respectively.
We assume, without loss of generality, $(\hb_i, \sigma_i)$ is the sufficient statistic derived from $\Xv_i$.
Let $\beta_i$ denote the true latent effect for the corresponding experiment and assume the grand effect in the population all (hypothetical) experiments is represented by $\bbar$.
Abstracting away from the details of the experiment design and data analysis, we consider the following simple probabilistic generative model:
\begin{equation} \label{meta.model}
\begin{aligned}
  & \bbar \sim {\rm N} (0, \omega^2) \\
  & \beta_i = \bbar + \eta_i,  ~ \eta_i \sim {\rm N}(0, \phi^2) \\
  & \hb_i = \beta_i + e_i, ~ e_i \sim {\rm N}(0, \sigma_i^2)
\end{aligned}
\end{equation}
Both $\eta_i$'s and $e_i$'s are assumed mutually independent across experiments.

To formulate a reproducible model, we apply the DC criterion to constrain the variability of $\beta_i$. Specifically, we parametrize $\gamma = \frac{\phi^2}{\phi^2 + \omega^2}$ and note that
  \begin{equation}\label{meta.dc}
%   \begin{aligned}
    p(\gamma) = \Pr \left( \, \sgn (\beta_{i}) = \sgn(\bar \beta)  \, \mid \, \gamma \right) = \sqrt{\frac{2}{\pi}} \, \int_{0}^{+\infty}\Phi\left(\sqrt{\frac{1-\gamma}{\gamma}} \, \xi\right)\, e^{-\frac{\xi^2}{2}}\, d \xi, 
%   & \mbox{and} \\
%   & \Pr \big( \, \sgn (\beta_{i}) = \sgn(\beta_{j})  \, \mid \, \gamma \big) = p(\gamma)^2 + \big(1- p(\gamma)\big)^2,~~~ \forall i \ne j.
%   \end{aligned}
\end{equation}
where $\Phi(\cdot)$ denotes the CDF of the standard normal distribution.
Thus, we can define a level of acceptable variability through the sign-consistency probability, $p(\gamma)$, under the DC principle \cite{Zhao2020}.

To complete a reference reproducible model covering diverse practical cases, we consider a grid of $(\omega^2, \gamma)$ from a Cartesian product $\Omega \times \Gamma$.
Each grid value is assigned equal prior probability, $\pi = 1/(|\Omega| \cdot |\Gamma|)$.
As the heterogeneity parameter $\gamma$ is of primary interest for inference and the effect size parameter $\omega$ is regarded as a nuisance parameter, we consider $\Gamma = \{\gamma: p(\gamma) = 1.00,\, 0.99,\, 0.975, \, 0.95 \}$ and construct $\Omega$ adaptively according to observed data and various application scenarios, by default. (See Appendix \ref{prior.spec} for details).
The default construction of $\Gamma$ specifies a lower bound for the sign-consistency probability (i.e., 0.95) but also allows more stringent criteria.
The discrete construction for the overall hyperparameter space results in a finite mixture model as the reference reproducible model. Thus, it provides not only modeling flexibility and comprehensiveness but also computational convenience for inference procedures.

For inference,  $\hat \bv =(\hb_1,...,\hb_m)$ are assumed observed. Without loss of generality in demonstrating the proposed methods, we also consider the true standard errors $\sigv =(\sigma_1,..., \sigma_m)$ are known or accurately measured.
Additionally, we illustrate the applications in the two-group scenario by considering two experiments ($m=2$), which are explicitly labeled as original and replication, respectively.

\subsection{Prior Predictive Checking  (for Two-group Scenario) }

The replication assessment in the two-group scenario requires explicit conditioning on the original experiment result.
It naturally leads to a predictive checking procedure in the Bayesian framework.
Specifically, a Bayesian predictive distribution is computed conditional on the original experiment and a reference reproducible model (denoted by $\Mr$), namely, $P(\Xv \mid \Xv_{\rm orig},\, \Mr)$.
A $p$-value is subsequently evaluated to quantify the probability of obtaining results at least as extreme as the observed replication data, assuming the predictive distribution.
A small $p$-value indicates poor model fitting by the replication data to the predictive distribution constructed from the original data and raises attention to potential irreproducibility.
Henceforth, we call the $p$-values derived from this procedure as the {\em prior-predictive replication $p$-values} (prior-PRPs).
Let $T(\Xv)$ denote a pre-defined test statistic, the one-sided version of the prior-predictive replication $p$-value is given by
\begin{equation}
   p_{\rm prior} := \Pr( \,T(\Xv) \ge T(\Xv_{\rm rep}) \mid \Xv_{\rm orig}, \Xv_{\rm rep}, {\rm M_R} \, );
\end{equation}
and correspondingly, the two-sided version is defined by
\begin{equation}
  \begin{aligned}
    p_{\rm prior} :=  2 \, \min \Big \{ &\Pr( \,T(\Xv) \ge T(\Xv_{\rm rep}) \mid \Xv_{\rm orig}, \Xv_{\rm rep}, {\rm M_R} \,)\,, \\
    & \Pr( \,T(\Xv) \le T(\Xv_{\rm rep}) \mid \Xv_{\rm orig}, \Xv_{\rm rep}, {\rm M_R} \,)\, \Big \}
 \end{aligned}
\end{equation}
%Our notations for $p$-values follow \cite{Meng1994} and \cite{Gelman}.

In the proposed reference reproducible model applying the DC criterion, $T(\Xv_{\rm orig}) = \hb_{\rm orig}$ and $T(\Xv_{\rm rep}) = \hb_{\rm rep}$ are natural test statistics.
Under the mixture prior on $(\omega, \gamma)$, the required posterior and predictive distributions based on the original data are analytically available for evaluating $p$-values (Appendix \ref{prior-prp.derive}).

The proposed procedure is considered as Bayesian prior predictive checking regarding the analysis of the replication data.
In the setting of the two-group scenario and following the principle of sequential Bayesian updating, the natural prior for estimating $\bbar$ (or $\br$ ) from the replication data is
$P(\bbar \mid \hbo, \Mr)$, which is the posterior from analyzing the original experiment.
It is important to note that, under $\Mr$, the original experiment can be highly informative on the magnitude of $\bbar$ and $\omega$, but it lacks information to alter the prior assumption on the heterogeneity parameter $\gamma$.
From this perspective, the replication data is used to examine the prior reproducibility assumption implementing the DC criterion.
To improve the interpretability of the proposed approach, we explicitly set the grid values of $\omega$ adaptive to the original data (Appendix \ref{prior.spec}).

The prior-predictive replication $p$-values are $u$-values \cite{Gelman2005, Gelman2013} defined by the following proposition.

\begin{proposition}
  Prior-predictive replication $p$-values are uniformly distributed on $[0,1]$ under the assumed reference reproducible model.
\end{proposition}
\begin{proof}
 Appendix \ref{prop1.proof}
\end{proof}

\noindent {\em Remark.} The uniform distribution is established under the repeated sampling of both $\Xv_{\rm orig}$ and $\Xv_{\rm rep}$. The exact uniformity is attained only under the {\em true} reference model.

The theoretical uniformity of the prior-predictive replication $p$-value is most helpful for benchmarking calibrations and evaluating sensitivity under a simulation setting when ground truths are known.
In practice, it provides a useful reference for simultaneous inspections of multiple sets of replication experiments. For example, in RP:P, $\sim 100$ pairs of original and replication experiments are selected to survey the replicability in psychology research.
If there is no {\em systematic} irreproducibility, an empirical distribution of prior-predictive replication $p$-values pooling all studies should closely resemble a discrete uniform distribution.
However, if the empirical distribution severely deviates from the uniformity and indicates clear positive skewness, the overall quality of replicability in the investigated set is questionable.
More generally, particular patterns of deviations from uniformity are informative on the average heterogeneity in the examined set comparing to the reference model.

The two-sided prior-predictive replication $p$-values are connected to the predictive interval checking procedures in replication assessment \cite{Patil2016, Pawel2020}. Specifically, the interval-checking procedures construct a $(1-\alpha)\%$ predictive interval based on an assumed reproducible model and the original experiment. They then examine if the observed replication data fall into the predictive intervals.
Let $q_\alpha(T(\Xv) \mid \Xv_{\rm orig}, \Mr)$ denote the quantile function of the predictive distribution of $T(\Xv)$ based on the original experimental data, such that $\Pr \left( T(\Xv) \le q_\alpha(T(\Xv) \mid \Xv_{\rm orig}, \Mr) \right) = \alpha$.
The predictive interval, $\left(  q_{\alpha/2}(T(\Xv)\,\mid \, \Xv_{\rm orig}, \Mr )\,,\, q_{(1-\alpha/2)}(T(\Xv)\,\mid \, \Xv_{\rm orig}, \Mr ) \right)$, is known as a $(1-\alpha)$ central interval \cite{Thulin2014}.
The following proposition represents the equivalence of this specific form of interval-checking procedure and the model criticism approach using the two-sided prior-predictive $p$-values derived from the same predictive distribution.

\begin{proposition}
 $T(\Xv_{\rm rep}) \not\in \left ( q_{\alpha/2}(T(\Xv)\,\mid \, \Xv_{\rm orig}, \Mr )\,,\, q_{(1-\alpha/2)}(T(\Xv)\,\mid \, \Xv_{\rm orig}, \Mr )  \right) $, if and only if the corresponding two-sided $p_{\rm prior} < \alpha$.
\end{proposition}
\begin{proof}
 Appendix \ref{prop2.proof}
\end{proof}

\noindent {\em Remark.} The similar equivalence can be generalized to one-sided prior-predictive replication $p$-values and the corresponding one-sided predictive intervals. 

The connection described by Proposition 2 provides not only an alternative interpretation of the proposed $p$-values but also a way of visualizing replication assessment.
Additionally, it highlights the distinction in the usage of $p$-values between our proposed model criticism framework and the NHST, especially in a setting of multiple comparisons.
As the type I error is irrelevant in our context, there is no need to adjust the predefined significance threshold (e.g., by Bonferroni correction) when multiple sets of experiments are simultaneously examined.
Similarly, it seems illogical to modify a well-defined predictive interval (derived from the corresponding original experiment) just because of the co-existence of other sets of the experiments.

It is worth pointing out that, from the model criticism perspective, replication data falling within the corresponding predictive interval is not evidence {\em in favor of} replication success. 
This argument is based on Proposition 2 and the interpretation of relatively large $p$-values (Principle 6 of the ASA statement on $p$-values, \cite{ASA2016}).
Only the rejection region, defined by the complement of a predictive interval in Proposition 2, is informative for evidence {\em against} the reference replication model.
A useful corollary is that the length of a rejection region can be used to measure the (relative) informativeness of the procedure.

By default, we use $T(\Xv_{\rm orig}) = \hbo$, $T(\Xv_{\rm rep}) = \hbr$ and compute two-sided $p$-values for replicability assessment under the two-group scenario.
Its predictive distribution based on the reference model, $\Mr$, can be analytically derived (Appendix \ref{prior-prp.derive}).
The derivation shows that replication noise, characterized by $\sigma_{\rm rep}$, does not change the location information of the prior predictive distribution. 
However, the length of a predictive interval monotonically increases with respect to $\sigma_{\rm rep}$. Correspondingly, the length of the are of resulting rejection region is monotonically decreasing. That is, the procedure becomes less informative as replication noise increases, which satisfies the non-informativeness principle.

\subsection{Posterior Predictive Checking (for Exchangeable-group Scenario) }

In the exchangeable-group scenario, the replicability assessment is conditional on all observed experiments.
To this end, we fit the reference model using the data from all experiments and compute a Bayesian $p$-value \cite{Rubin1984, Meng1994, Gelman1996} based on the corresponding posterior predictive distribution, $P(\Xv \mid \Xv_1,...,\Xv_m, \Mr)$.
We refer to the Bayesian $p$-values derived from this procedure as the {\em posterior-predictive replication $p$-values} (posterior-PRPs).

A unique property of the posterior predictive checking is its usage of test quantities (also known as discrepancy variables).
A test quantity, denoted by $T(\Xv, \tv)$, is a function of both observed data ($\Xv$) and latent variables/hyperparameters ($\tv$) defined by a reference model.
In comparison, traditional test statistics are functions of only observed data ($\Xv$).
A one-sided posterior-predictive replication $p$-value is defined and computed by
\begin{equation} \label{posterior.pvalue}
  \begin{aligned}
   p_{\rm posterior} & := \Pr\Big( \, T(\Xv_1', \dots,\Xv_m', \tv) \ge T(\Xv_1, \dots,\Xv_m, \tv) \, \mid \, \Xv_1, \dots,\Xv_m, \Mr \Big) \\
       &~ =  \int_{\big(\tv, \Xv_1',...,\Xv_m'\big)} \mathlarger{\mathds{1}}\big( T(\Xv_1', \dots,\Xv_m', \tv) \ge T(\Xv_1, \dots,\Xv_m, \tv)\big) \\
       & ~~~~\cdot \left(\prod_{i=1}^m P(\Xv_i' \mid \tv )\right) P(\tv \mid \Xv_1,...,\Xv_m)\,\,d\tv\,d\Xv_1'\,\dots\, d\Xv_m'\,.
\end{aligned}
\end{equation}
We use a posterior sampling scheme, outlined in Algorithm \ref{post.sample.alg}, to approximate the exact $p$-values.
The relevant mathematical details in the algorithm are provided in Appendix \ref{posterior-prp.derive}.
\begin{algorithm}
  \caption{Computing posterior-predictive replication $p$-values}
  \label{post.sample.alg}
  \begin{algorithmic}[1]
  \Procedure{Approximating $p$-value by posterior sampling}{}
  \State compute posterior distribution $P(\tv \mid \Xv_1, \dots, \Xv_m, \Mr)$
  \State initialize counter $ l \gets 0$
  \For{  $k = 1 $ to $L$ }
  \State sample $\tv \sim P(\tv \mid \Xv_1, \dots, \Xv_m, \Mr)$
  \State independently sample $\Xv_i' \sim P(\Xv \mid \tv, \Mr)$ for $i=1,...,m$
  \State evaluate $T(\Xv_1,...,\Xv_m, \tv)$ and $T(\Xv_1',...,\Xv_m', \tv)$
  \State set $l \gets l + 1$ if $\, T(\Xv_1',...,\Xv_m', \tv) \ge T(\Xv_1,...,\Xv_m, \tv)$
  \EndFor
  \State \Return $p$-value $ = l/L$
\EndProcedure
\end{algorithmic}
\end{algorithm}

The interpretation of the posterior-PRPs is not different from the prior-PRPs: a small value indicates poor model fitting and suggests that the observed data are unlikely under the assumed reference model.
In our implementation, we set the hyperparameter, $\omega$, adaptively to the observed data (Appendix \ref{prior.spec}). Hence, the lack of model fit implies potential violations of the replicability assumption defined by the DC criterion.

The posterior-predictive replication $p$-values, as a special case of Bayesian $p$-values, are not necessarily uniformly distributed on $[0,1]$ under the assumed reference model and repeated sampling of $(\Xv_1,...,\Xv_m)$.
Meng \cite{Meng1994} shows that under the true reference model, a Bayesian $p$-value can be stochastically less variable than a uniform distribution on $[0, 1]$ but with the same mean ($1/2$).
This implies that extremely small Bayesian $p$-values are less likely under the true reference model than a frequentist (or prior-predictive replication) $p$-value.
As we have emphasized that the proposed model criticism approaches fundamentally differ from the Neyman-Pearson hypothesis testing and NHST, the non-uniformity property has little impact on interpreting results from evaluating an individual set of experiments.
This is because the Bayesian $p$-values are valid posterior probabilities, as Gelman points out in\cite{Gelman2005, Gelman2013}, and their usage in model criticism approaches is well-justified.
On the other hand, it is relevant to establish a benchmark if the goal is to jointly evaluate multiple sets of experiments (as we have discussed in the case of prior-PRPs).
In such a scenario, generating empirical distribution of the posterior-predictive replication $p$-values under the well-defined reference model for corresponding test quantities should suffice.

Motivated by Cochran's $Q$ statistic, we use a test quantity, $ T(\Xv, \tv) = T_{Q} \left(\hbv, (\bbar, \phi)\right)$, by default in our proposed reference reproducible model, i.e.,
\begin{equation}
   T_{Q} \left(\hbv, (\bbar, \phi) \right) = \sum_{i=1}^m w_i (\hb_i - \bbar)^2, \mbox{ where  } w_i = \frac{1}{\sigma_i^2 +\phi^2}
\end{equation}
Similar to Cochran's $Q$ statistic, this quantity measures the spread of the observed effects and quantifies the heterogeneity across $m$ participating experiments. But unlike the $Q$ statistic, which is computed under a fixed-effect assumption, $\phi^2$ is no longer constrained to be 0.

%%%%%%%%%%%%%%%%%%%%%%%%%%%%%%%%%%%%%%%%%%%

\section{Constructing Test Statistics and Quantities}

Although the proposed prior- and posterior-predictive checking procedures using the default test statistics/quantities serve the general purpose of replicability assessment, some applications attempt to investigate specific mechanisms of irreproducibility that lead to particular patterns in observed data.
To this end, the most effective way is to design and incorporate special-purpose test statistics/quantities into the proposed model criticism framework.

Take detection of publication bias as an illustrating example in two-group scenarios.
Because publication bias often leads to overestimating true effects in original experiments, the re-estimated effects in replication experiments tend to shrink noticeably towards 0.
To capture such particular directional shift of effect estimates, we consider the following test statistic,
\begin{equation}
  T_{\rm pb} = \frac{\hbr}{\hbo}.
\end{equation}
Stronger shrinkage effects lead $T_{\rm pb}$ further away from 1, regardless of the sign of $\hbo$.   
Thus, a one-sided prior-predictive replication $p$-value, i.e.,
\begin{equation}
\label{prior_prp.pb}
  p_{\rm prior,\,pb} := \Pr \left( T_{\rm pb} \le \frac{\hbr}{\hbo}\, \, \Bigl\lvert \, \,  \hbo, \hbr,\, \Mr \right)
\end{equation}
can be applied to detect patterns of publication bias.
The resulting procedure show much improved sensitivity to publication bias than the default statistic.

In the exchangeable-group scenario, detecting publication bias is a long-standing problem in meta-analysis and systematic review. The patterns of publication bias are traditionally investigated by detecting the asymmetry of funnel plots.
The inspection can be analytically carried out by Egger regression or other types of meta-regression procedures.
These approaches examine the correlation between the estimated effect, $\hat \beta_i$, and the corresponding standard error, $\sigma_i$, which characterizes a potential linear trend in funnel plots \cite{Sterne2005}. 
The implementation by Egger regression estimates the correlation in a weighted linear regression model, whose test statistic can be trivially replicated in the framework of posterior-PRPs. 
In the presence of genuine between-experiment heterogeneity, $\phi^2$, the theory requires to compute the correlation between $\hat \beta_i$ and $\sqrt{\sigma_i^2 + \phi^2}$.
The prevailing approach is to plug in an unconstrained point estimate of $\phi^2$. 
However, the accuracy of the point estimate is questionable, especially when the number of available experiments ($m$) is limited, and the solution also ignores its uncertainty. 
The proposed Bayesian model criticism approach naturally addresses this issue by explicitly sampling reference model-defined $\phi^2$ from the posterior distribution in computing the test quantity.

In summary, constructing test statistics for computing prior-PRPs is not fundamentally different from the common practice of significance testing, whereas the test quantities in posterior-predicting replication $p$-value computation present some unique methodological and practical advantages.
First, Bayesian $p$-values allow direct incorporation of hyperparameters and latent variables (i.e., $\tv$), which are typically considered nuisance parameters in constructing traditional test statistics.
The Bayesian procedure provides a principled way to account for their uncertainty without restricting the test quantities to be pivotal.
This feature allows for adopting more rational reference models permitting reasonable between-experiment heterogeneity in the context of replicability assessment. 
Second, the ``null'' distribution of the test quantity does not need to be pre-defined or pre-computed.
This flexibility allows practitioners to emphasize context-dependent information in constructing relevant test quantities rather than focus on their statistical properties.

\section{Numerical Illustrations}

We use simulation studies to mimic two common phenomenons that can lead to irreproducible results, namely, batch effects and publication bias. We examine the behaviors of the proposed statistical methods in different application scenarios. 

\subsection{Simulations of Batch Effects}

In association analysis, batch effects refer to unaccounted experimental factors causing spurious correlations. 
We consider a balanced case-control analysis of a quantitative trait with 200 individuals per experiment. 
For each experiment contaminated by batch effects, we simulate binary batch labels for all samples by permuting their case-control labels. 
On average, the correlation between the batch labels and the case-control status is 0.7. 
The continuous outcome of interest is generated from a linear regression model.

\subsubsection{Batch Effect Detection in Two-group Scenario}

To create a two-group scenario, we simulate two experiments following the general descriptions above. 
More specifically, the data from the original experiment is batch contaminated, where the batch effect is drawn from the distribution, ${\rm N} (0, \eta^2)$. 
The replication experiment is generated free of batch effect. 
The true case-control associations are fixed at 0.5 for both original and replication experiments, and the residual errors for each individual are independently sampled from the standard normal distribution. 
We vary $\eta$ from 0 to 1 to create different levels of contamination.
For each experiment, regardless of its contamination status, we fit a simple linear model by regressing the simulated outcome on the corresponding case-control status. 
The resulting association effect estimate and the corresponding standard error are used as summary statistics for input. We repeat the data generation and analysis procedures 5,000 times for each experimented $\eta$ value.  

We compute the prior-PRP for each simulated dataset, assuming the default reproducible model. 
A simulated dataset is flagged if its corresponding prior-predictive $p$-value is smaller than the commonly used significance threshold 0.05.
We define {\em sensitivity} as the proportion of flagged datasets for each experimented non-zero $\eta$ value.   

\begin{figure}[H]
\centering
    \begin{subfigure}[b]{0.45\textwidth}
        \caption{$\eta = 0$}
        \includegraphics[width=\textwidth]{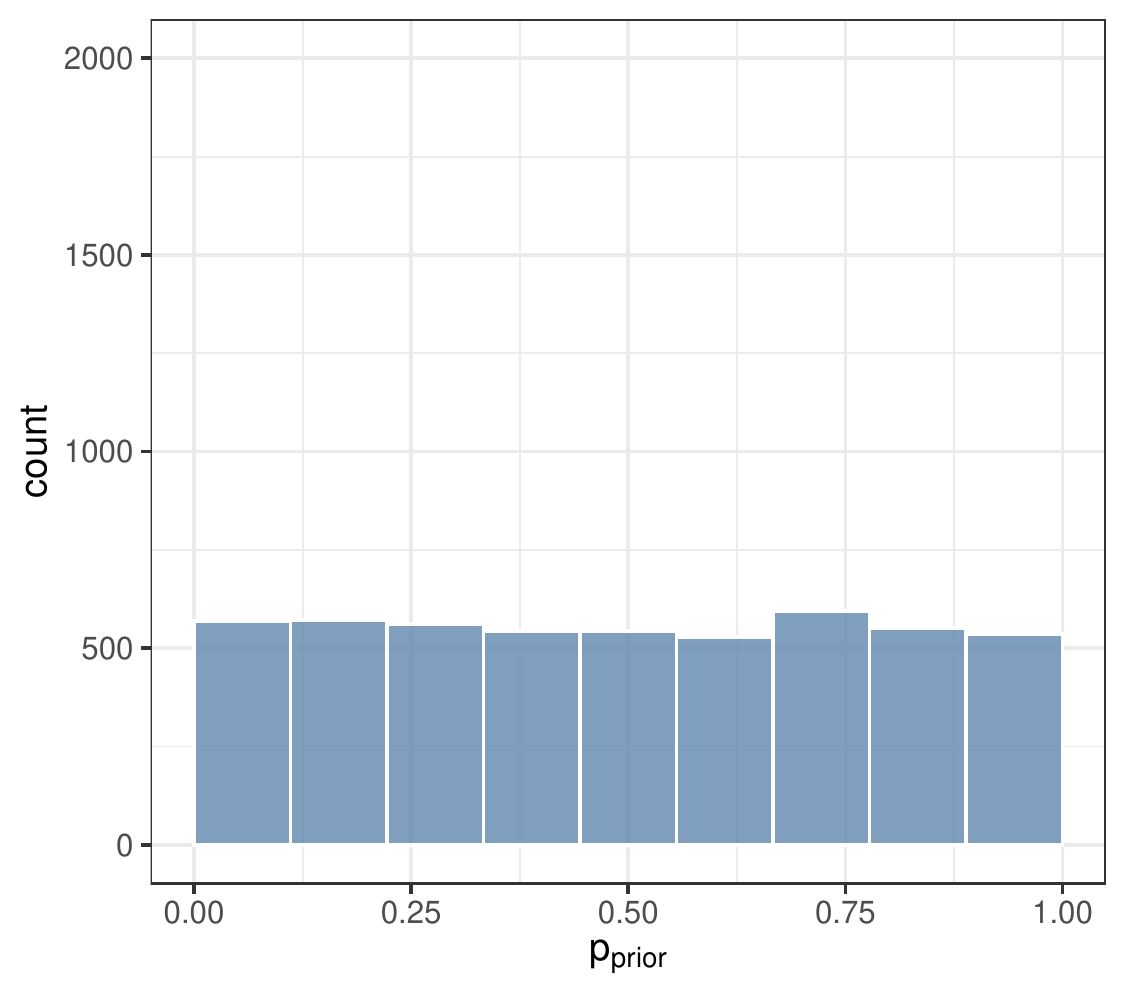}
        \label{bb0}
    \end{subfigure}
    \qquad
      \begin{subfigure}[b]{0.45\textwidth}
         \caption{$\eta = 0.6$}
        \includegraphics[width=\textwidth]{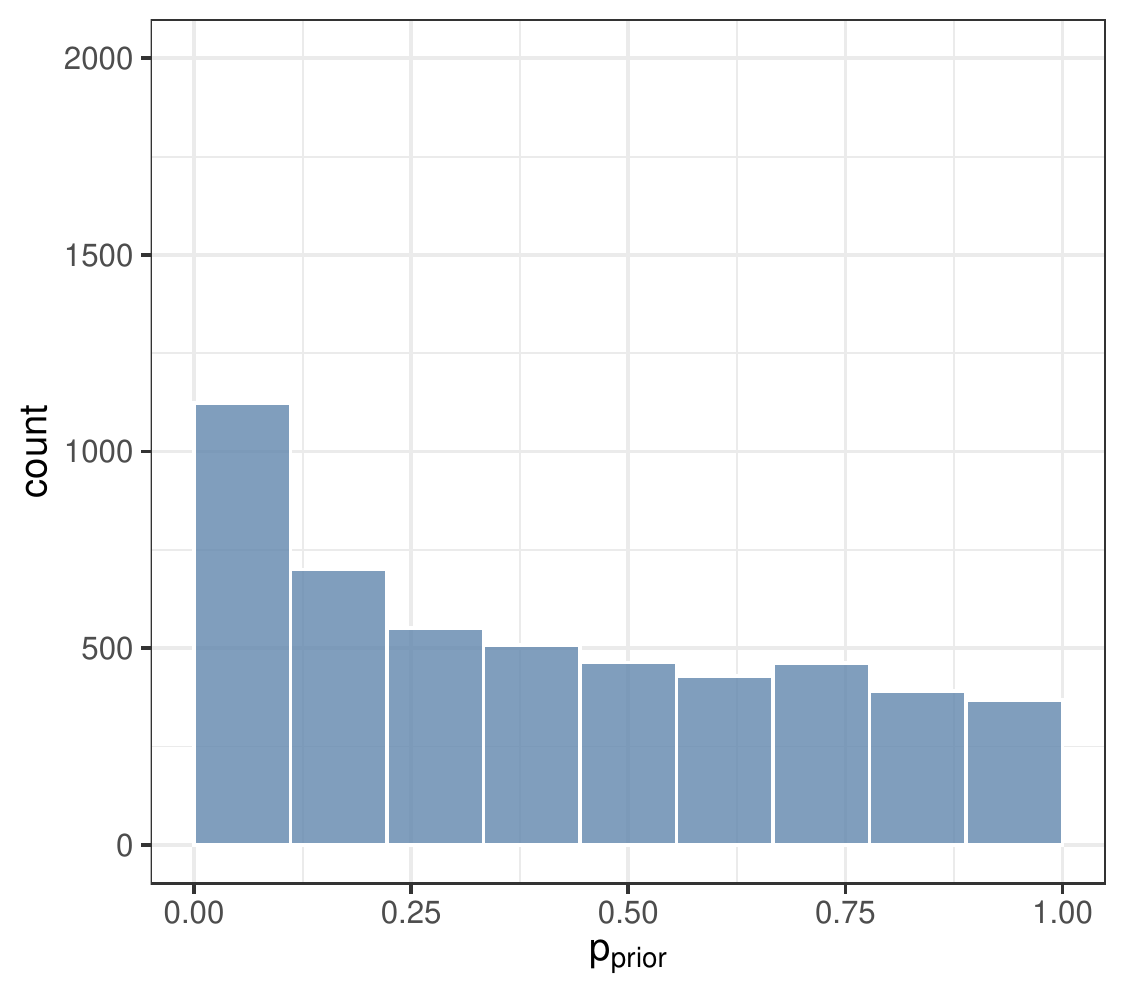}
        \label{bb06}
    \end{subfigure}
    \\
      \begin{subfigure}[b]{0.45\textwidth}
       \caption{$\eta = 1$}
        \includegraphics[width=\textwidth]{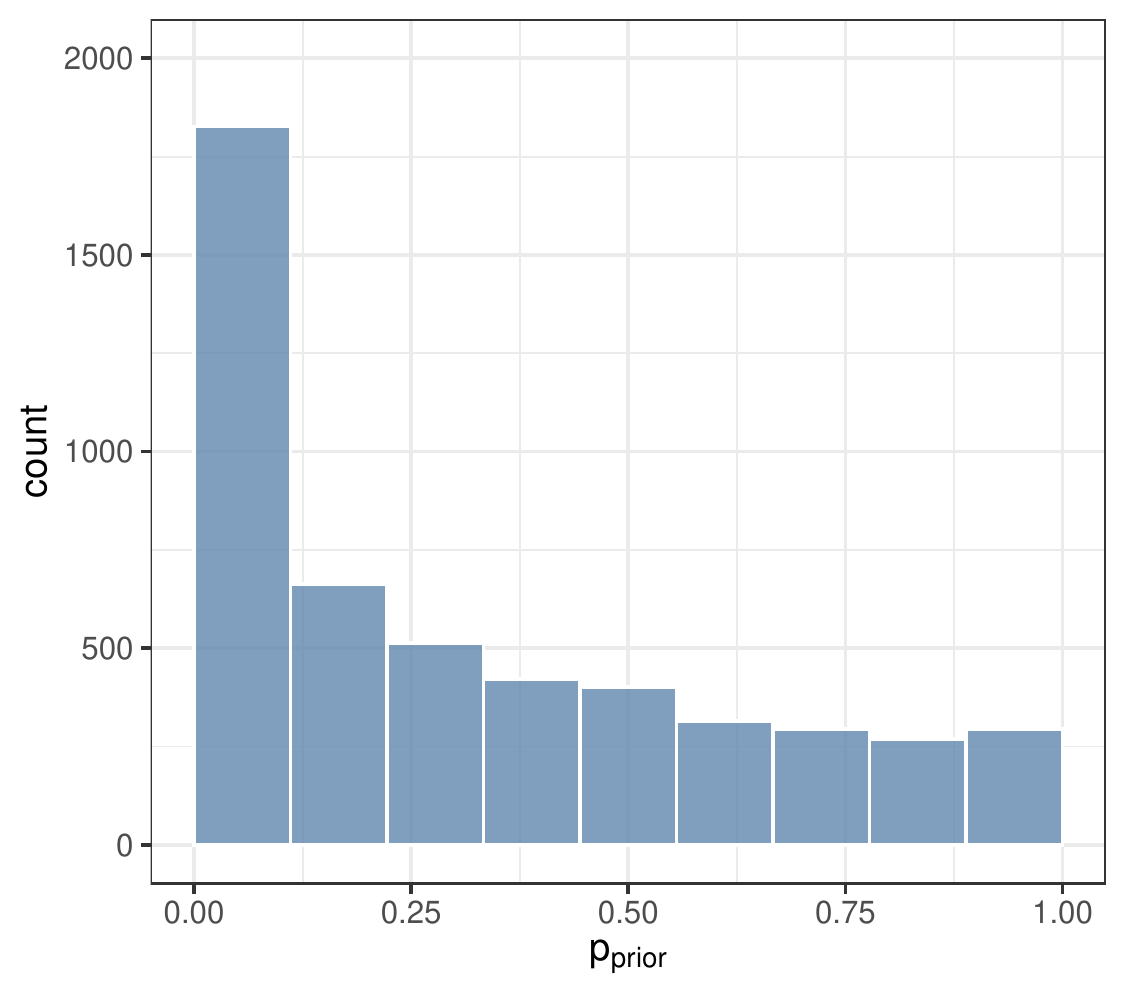}
        \label{bb1}
    \end{subfigure}
    \qquad
       \begin{subfigure}[b]{0.45\textwidth}
        \caption{Sensitivity}
        \includegraphics[width=\textwidth]{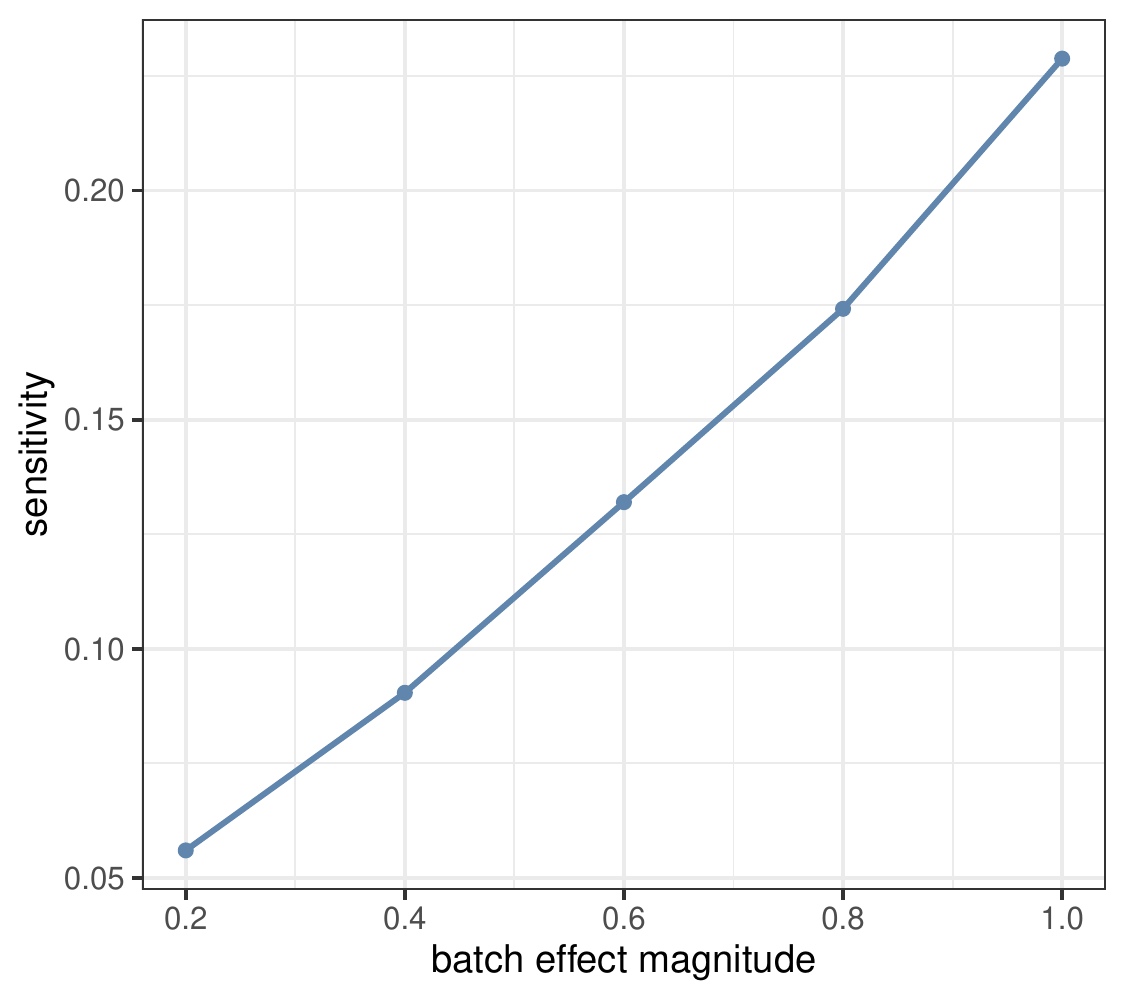}
        \label{sens}
    \end{subfigure}
    \caption{{\bf Batch effect detection in two-group scenario.} Panels {\bf a, b}, and {\bf c} present the histograms of the two-sided prior-PRPs from different simulation settings, where the magnitude of batch effects in the original experiments varies from $\eta = 0$ (i.e., no batch effect), $\eta = 0.6$, and to $\eta = 1.0$. Panel {\bf d} displays the relationship between the sensitivity (percentage of contaminated experiments detected by the replicability analysis) and the experimented batch effect magnitude.} \label{batch2gp}
\end{figure}

The simulation results are summarized in Figure \ref{batch2gp}. The empirical distribution of the prior-predictive $p$-values under $\eta = 0$ resembles a discrete uniform distribution, suggesting the desired calibration under the reference model.  
As expected, the sensitivity in detecting irreproducibility increases as the magnitude of batch effects increase.
The empirical $p$-value distributions also show increased positive skewness as $\eta$ increases.    

\begin{figure}[H]
\centering
    \begin{subfigure}[b]{0.45\textwidth}
        \caption{$\sigma = 5$}
        \includegraphics[width=\textwidth]{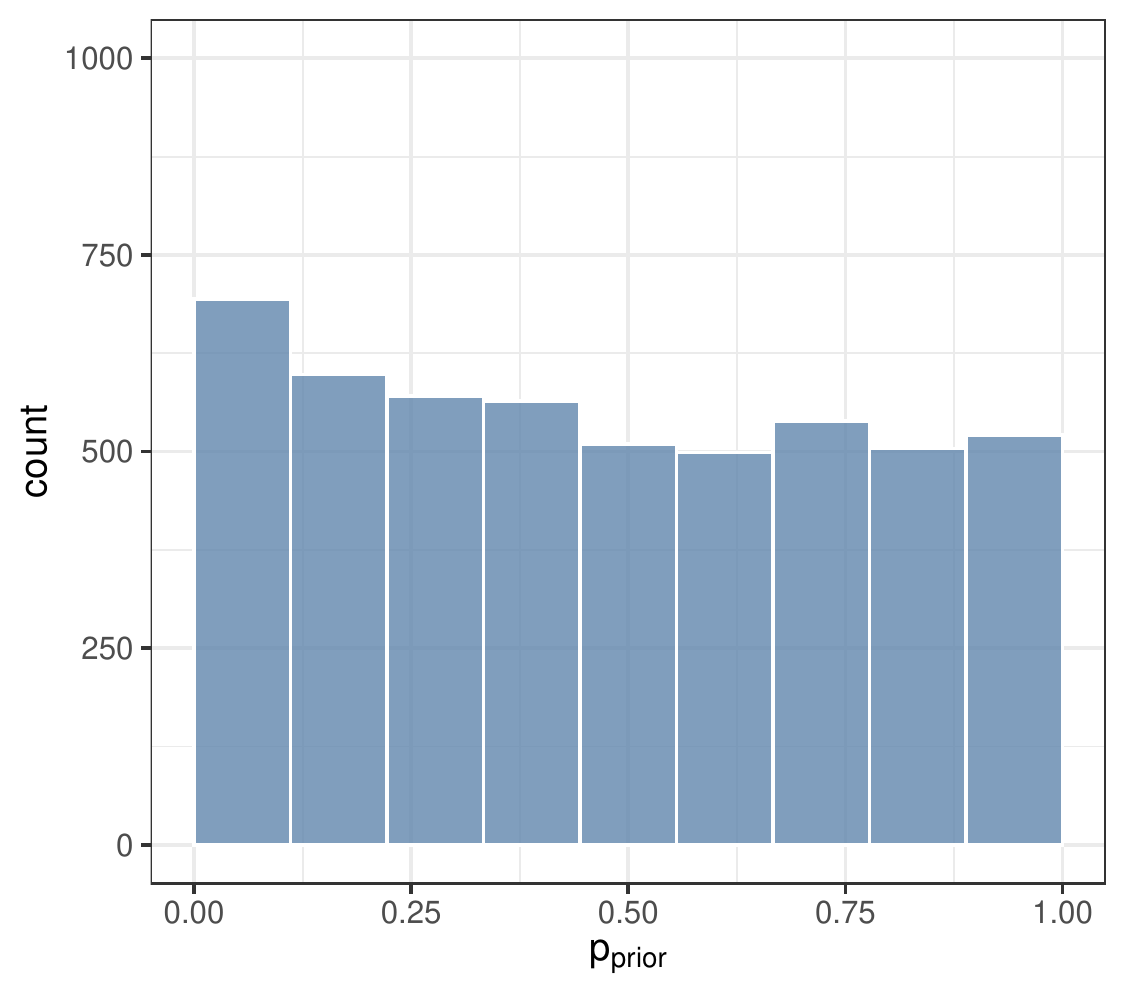}
        \label{e5}
    \end{subfigure}
    \qquad
      \begin{subfigure}[b]{0.45\textwidth}
        \caption{$\sigma = 10$}
        \includegraphics[width=\textwidth]{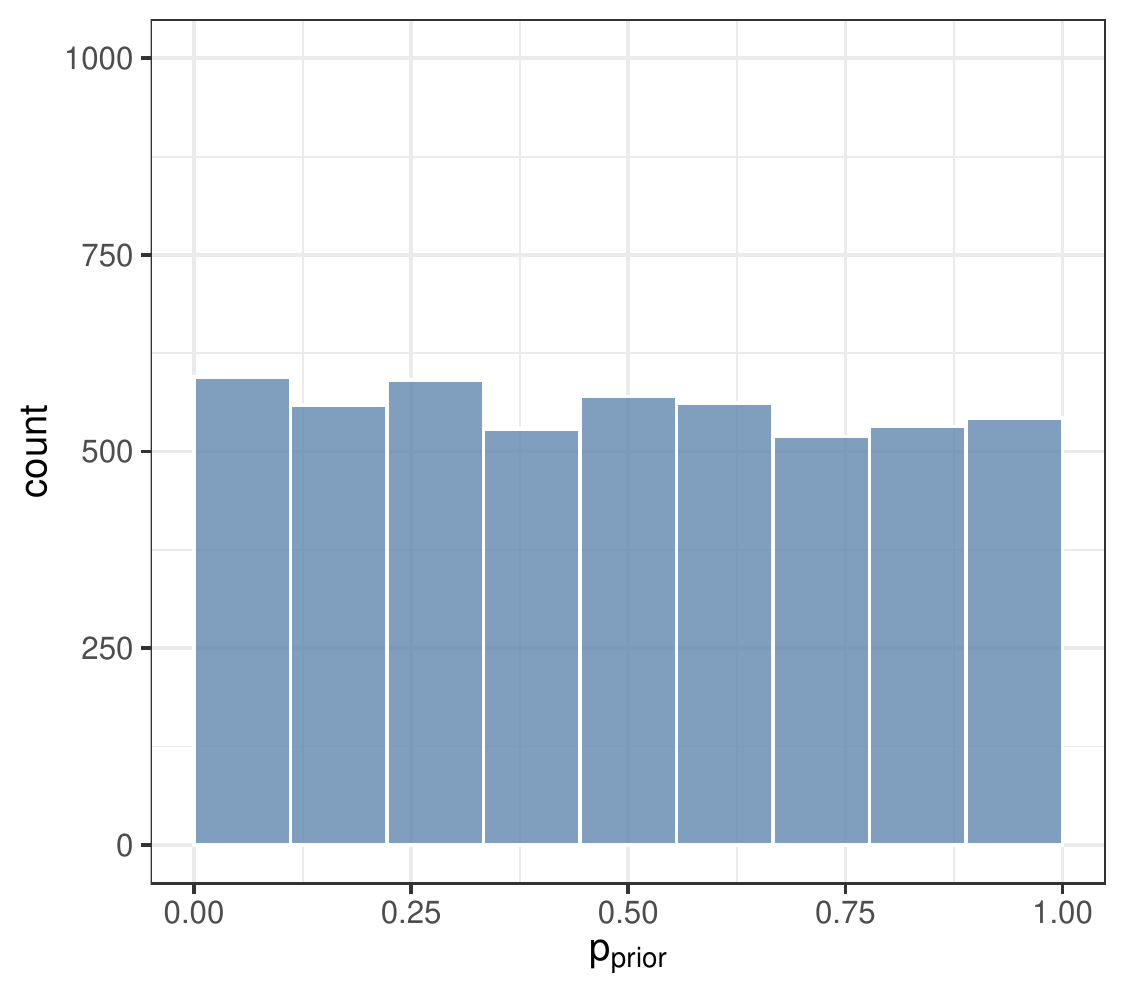}
        \label{e10}
    \end{subfigure}
\caption{{\bf Impacts of noisy replications on prior-PRPs~} The original experiments are simulated with batch effect magnitude $\eta = 1$. The data from replication experiments are not batch contaminated but generated with increasing levels of residual errors, i.e., $\sigma=5$ (Panel {\bf a}) and $\sigma = 10$ (Panel {\bf b}). Note that $\sigma_{\rm rep}$ is proportional to $\sigma$, and the baseline comparison is the panel {\bf c} of Figure \ref{batch2gp}, where $\eta = 1$ and $\sigma = 1$. }\label{noninfo}
 \end{figure}

To demonstrate the non-informativeness principle, we simulate additional datasets with $\eta=1$ but increased residual errors ($\sigma = 5, 10$) in replication experiments. 
Note that $\sigma_{\rm rep}$ is proportional to $\sigma$.
The resulting empirical distributions of prior-predictive replication $p$-values display a clear trend of reduced skewness (Figure \ref{noninfo}), suggesting decreased sensitivity.

\subsubsection{Batch Effect Detection in Exchangeable-group Scenario}

We adopt a similar scheme to simulate batch effects and generate observed summary statistics for the exchangeable-group scenario. 
For each dataset, we consider 5 experiments with 2 out of 5 experiments batch contaminated. 
Additionally, we introduce a low-level heterogeneity for experiment-specific association effects, such that the sign-consistency probability $\approx 0.96$ across experiments.

\begin{figure}[H]
\centering
    \begin{subfigure}[b]{0.35\textwidth}
            \caption{$ \eta= 0$}
        \includegraphics[width=\textwidth]{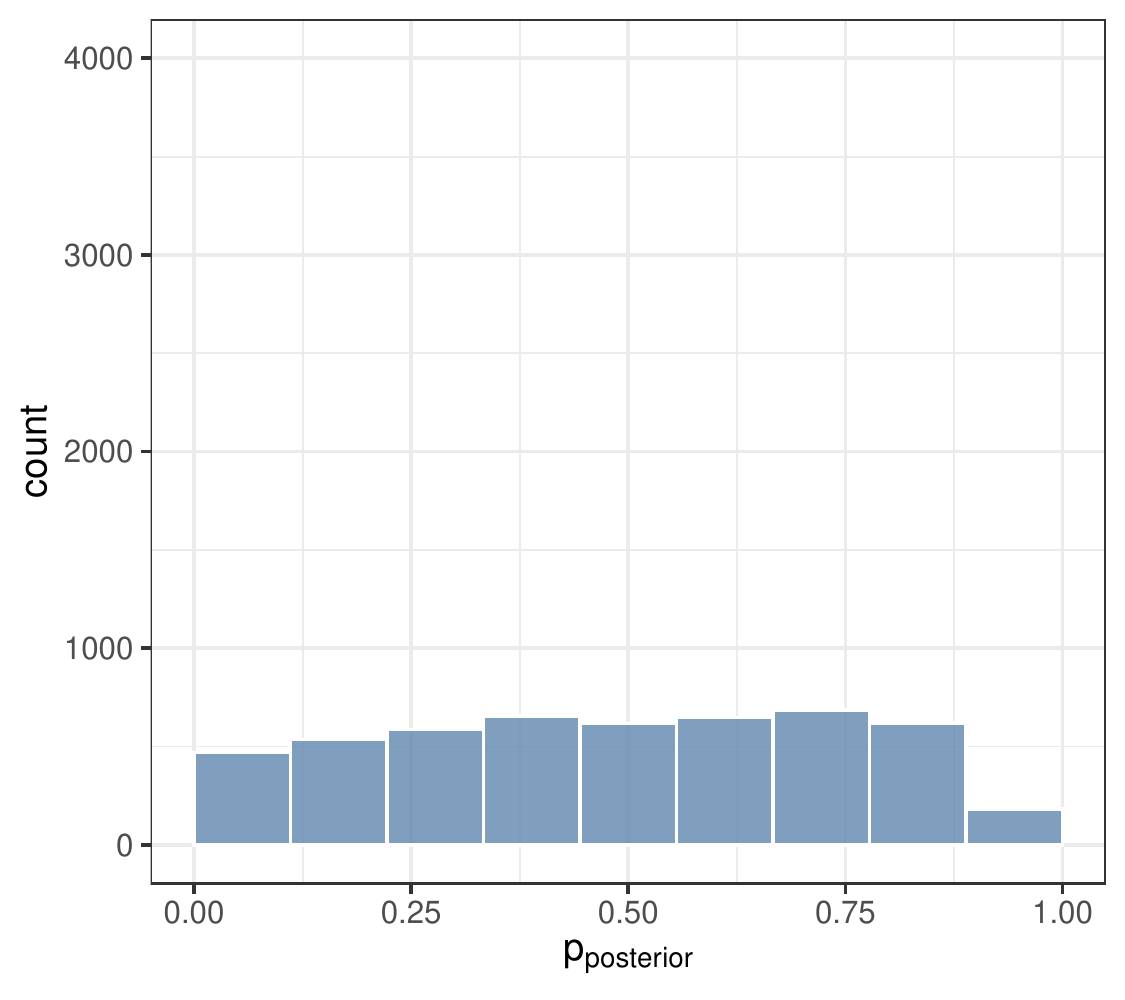}
    \end{subfigure}
    \qquad
      \begin{subfigure}[b]{0.35\textwidth}
              \caption{$\eta= 0.6$}
        \includegraphics[width=\textwidth]{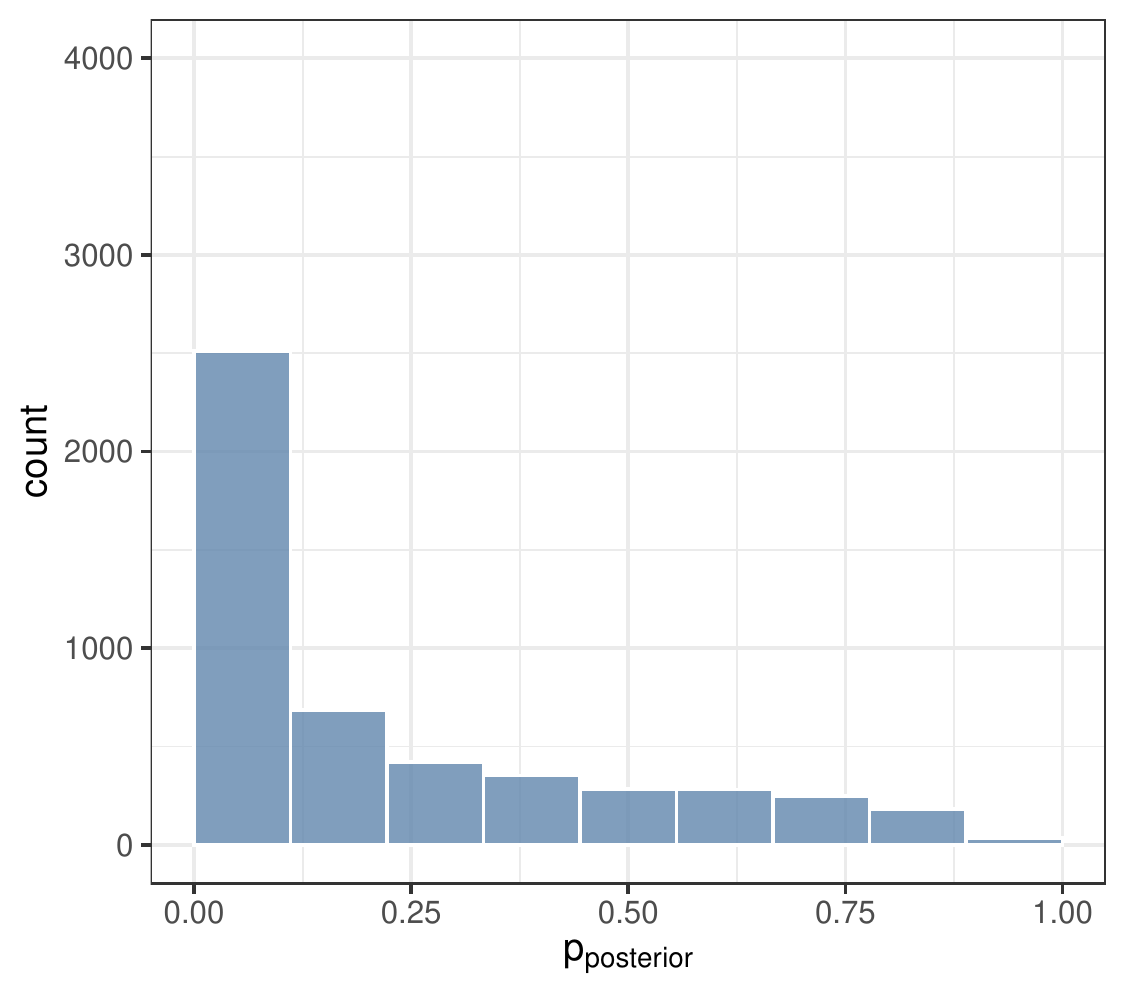}
    \end{subfigure}
    \\
      \begin{subfigure}[b]{0.35\textwidth}
              \caption{$\eta = 1$}
        \includegraphics[width=\textwidth]{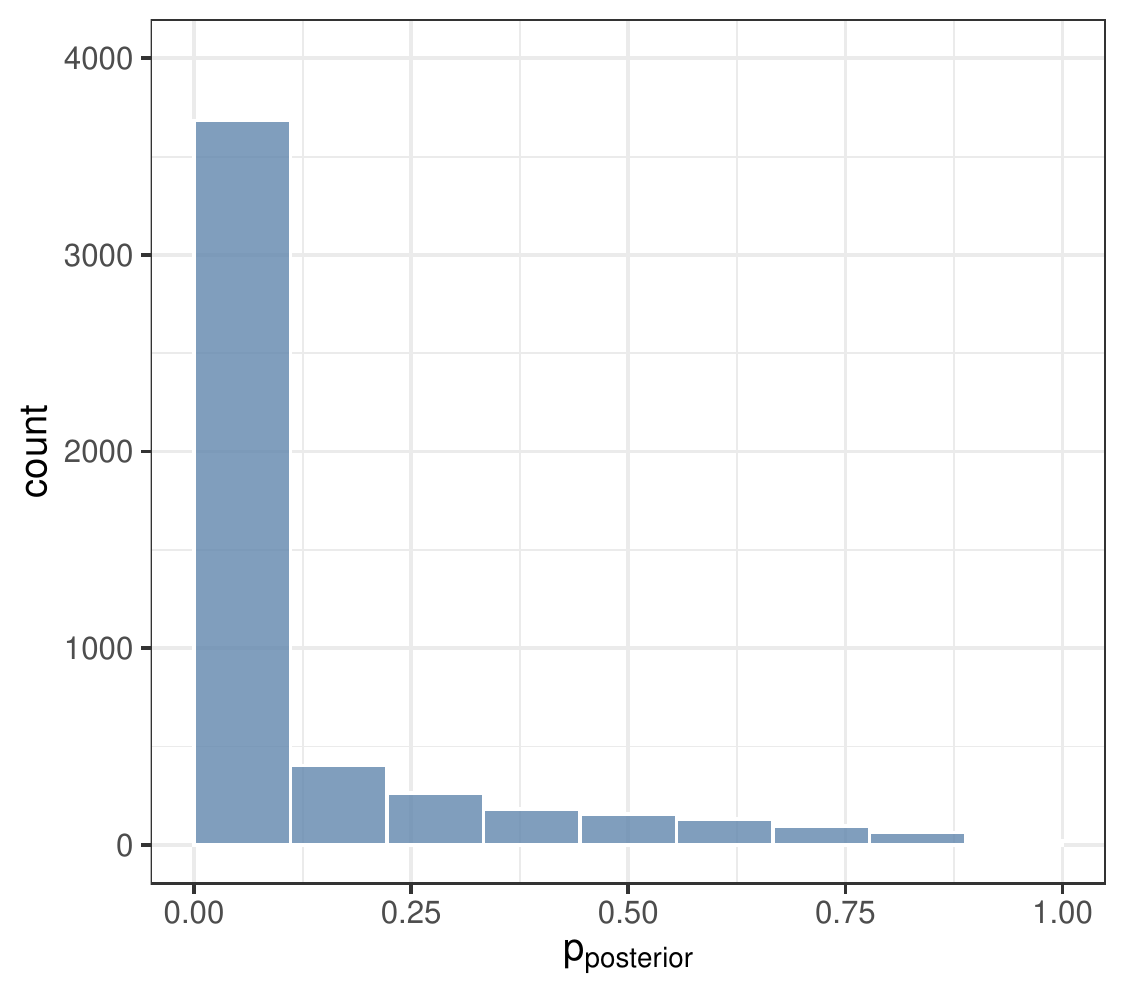}
    \end{subfigure}
        \qquad
       \begin{subfigure}[b]{0.35\textwidth}
               \caption{Sensitivity}
        \includegraphics[width=\textwidth]{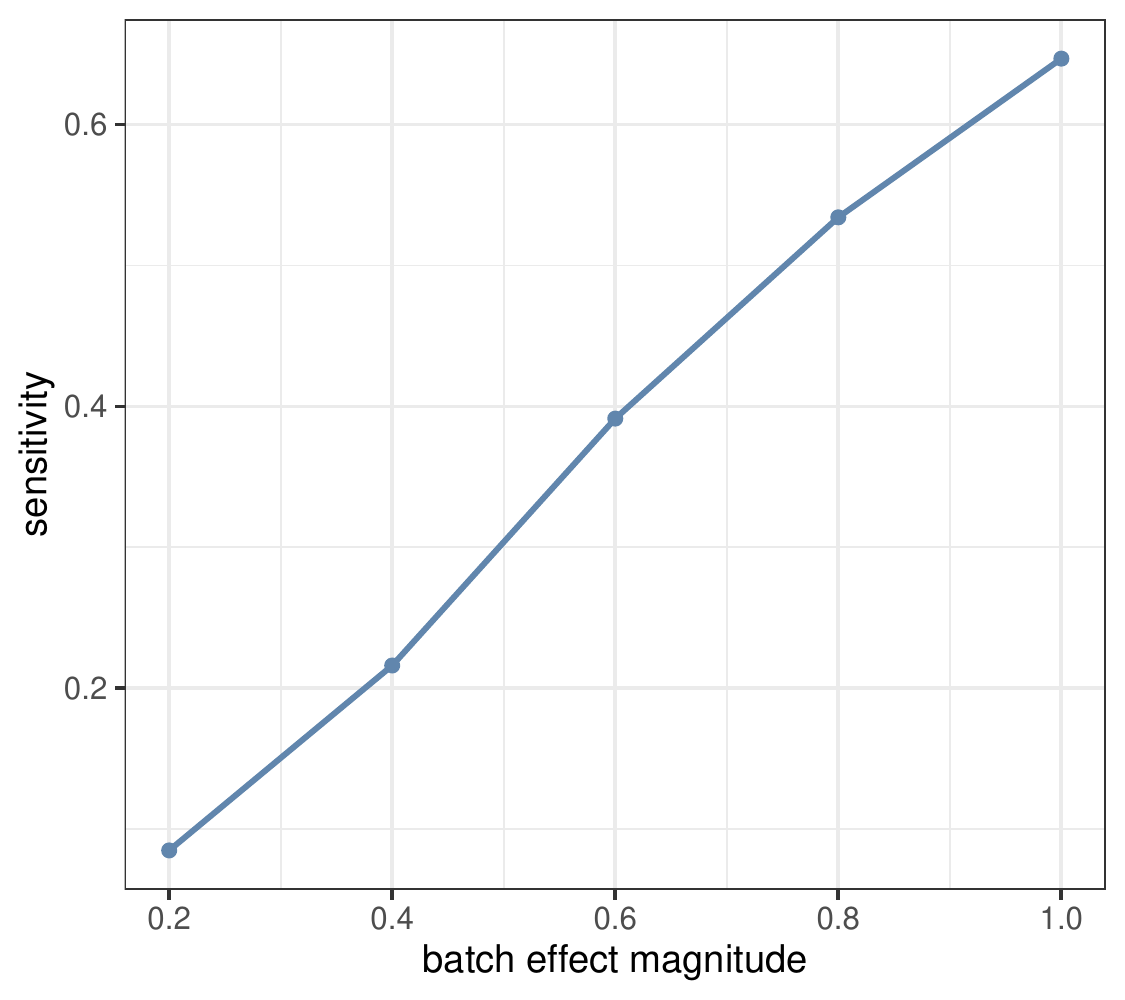}
    \end{subfigure}
    \caption{{\bf Batch effect detection in exchangeable-group scenario ~} In each dataset, we simulate five experiments and two are contaminated by batch effects with magnitude $\eta = 0$ (Panel {\bf a}, no batch effect), $\eta = 0.6$ (Panel {\bf b}), and $\eta = 1$ (Panel {\bf c}). Panel {\bf d} shows increased sensitivity of the posterior PRPs with respect to the increased batch effect magnitude. }\label{batchmultigrp}
\end{figure}

We compute a posterior-predictive replication $p$-value for each simulated dataset using the default Q test quantity and assuming the default mixture reference model.
Same as in the two-group scenario, increased sensitivity in detecting batch effects is observed as the magnitude of the batch effects ($\eta$) increases (Figure \ref{batchmultigrp}).  
Although the distributional properties of posterior-PRPs are not completely understood under the reference model, their empirical distributions under the simulation setting $\eta \ne 0$ all show clear positive skewness, indicating excessive small $p$-values.

\begin{figure}[H]
\centering

    \begin{subfigure}[b]{0.35\textwidth}
        \caption{}
        \includegraphics[width=\textwidth]{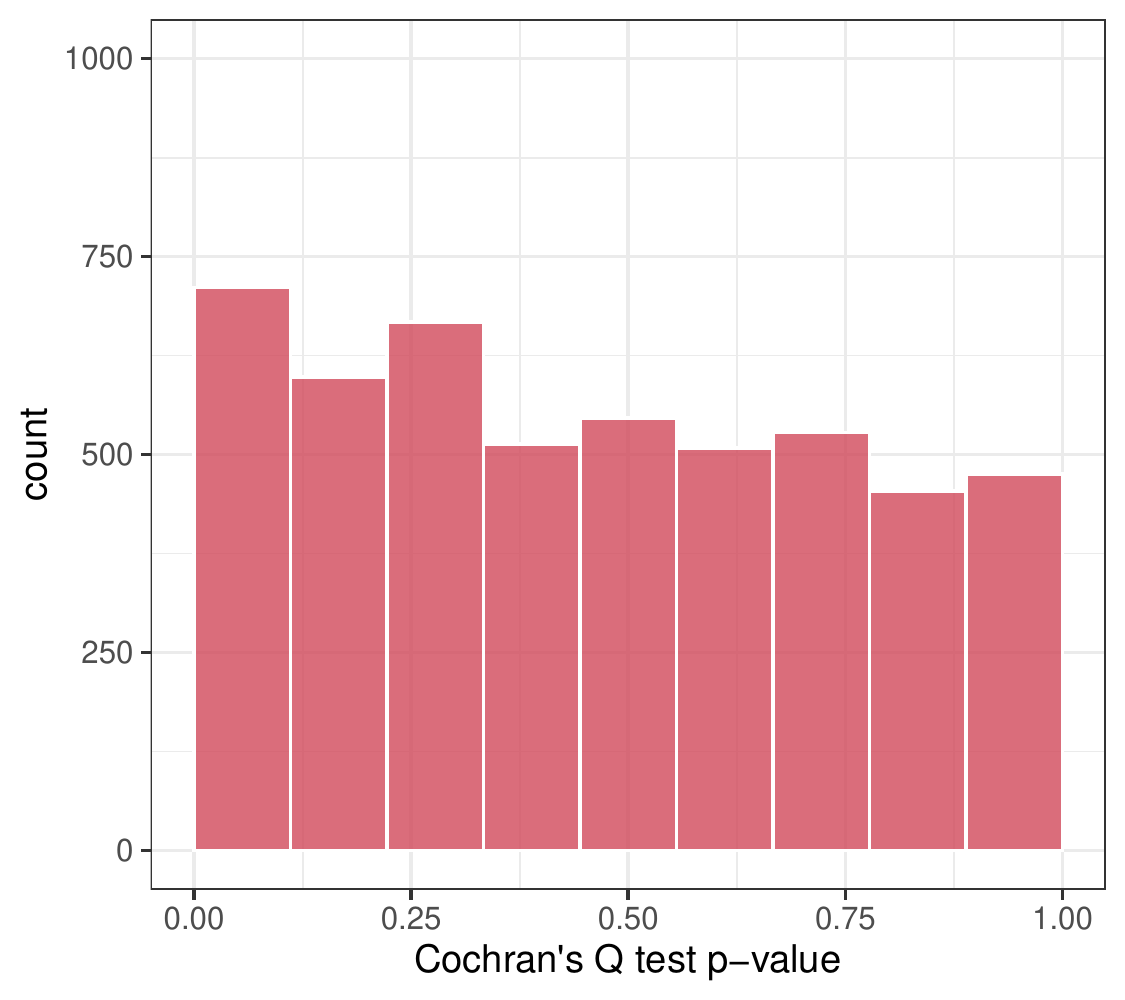}
    \end{subfigure}
    \qquad
    \begin{subfigure}[b]{0.35\textwidth}
        \caption{}
        \includegraphics[width=\textwidth]{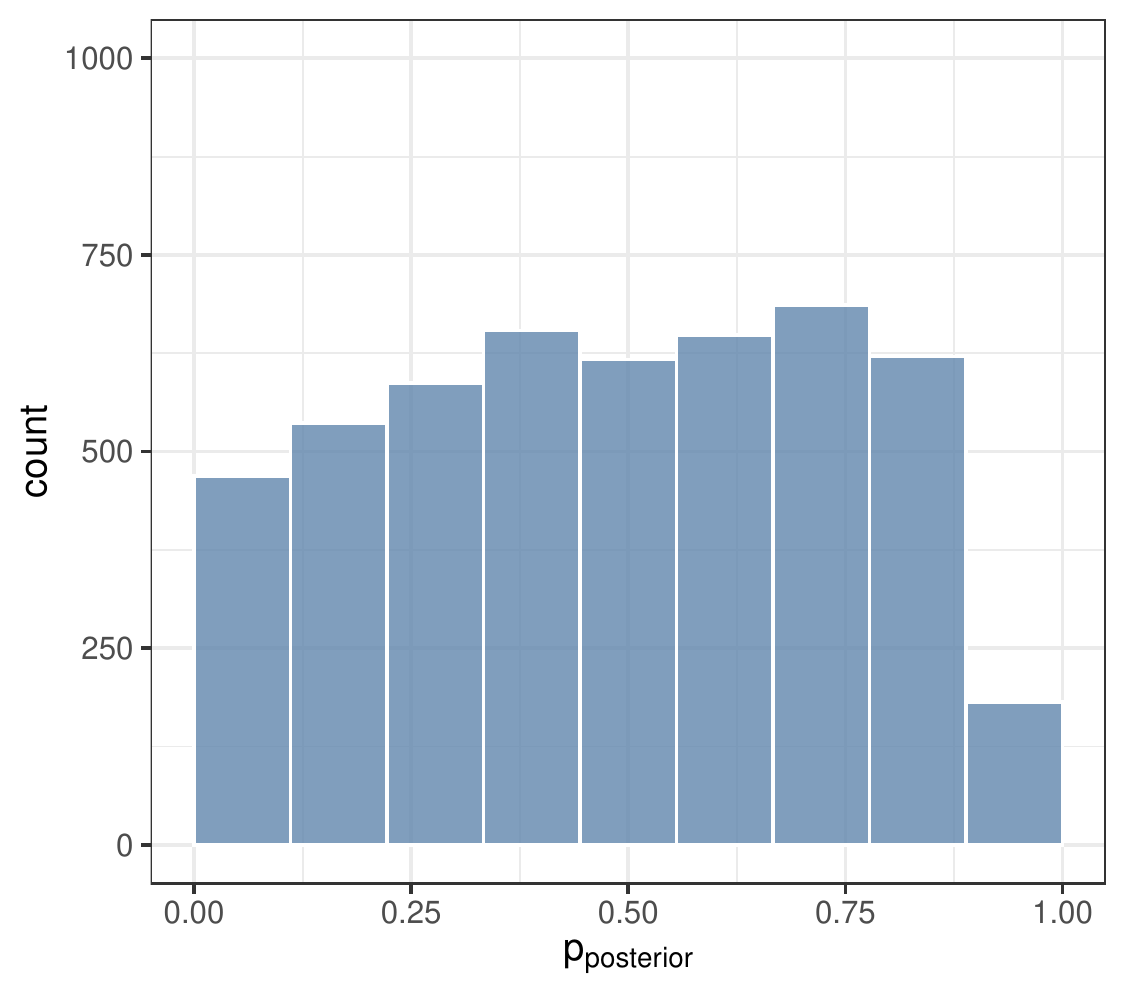}
    \end{subfigure}
    \\
      \begin{subfigure}[b]{0.35\textwidth}
        \caption{}
        \includegraphics[width=\textwidth]{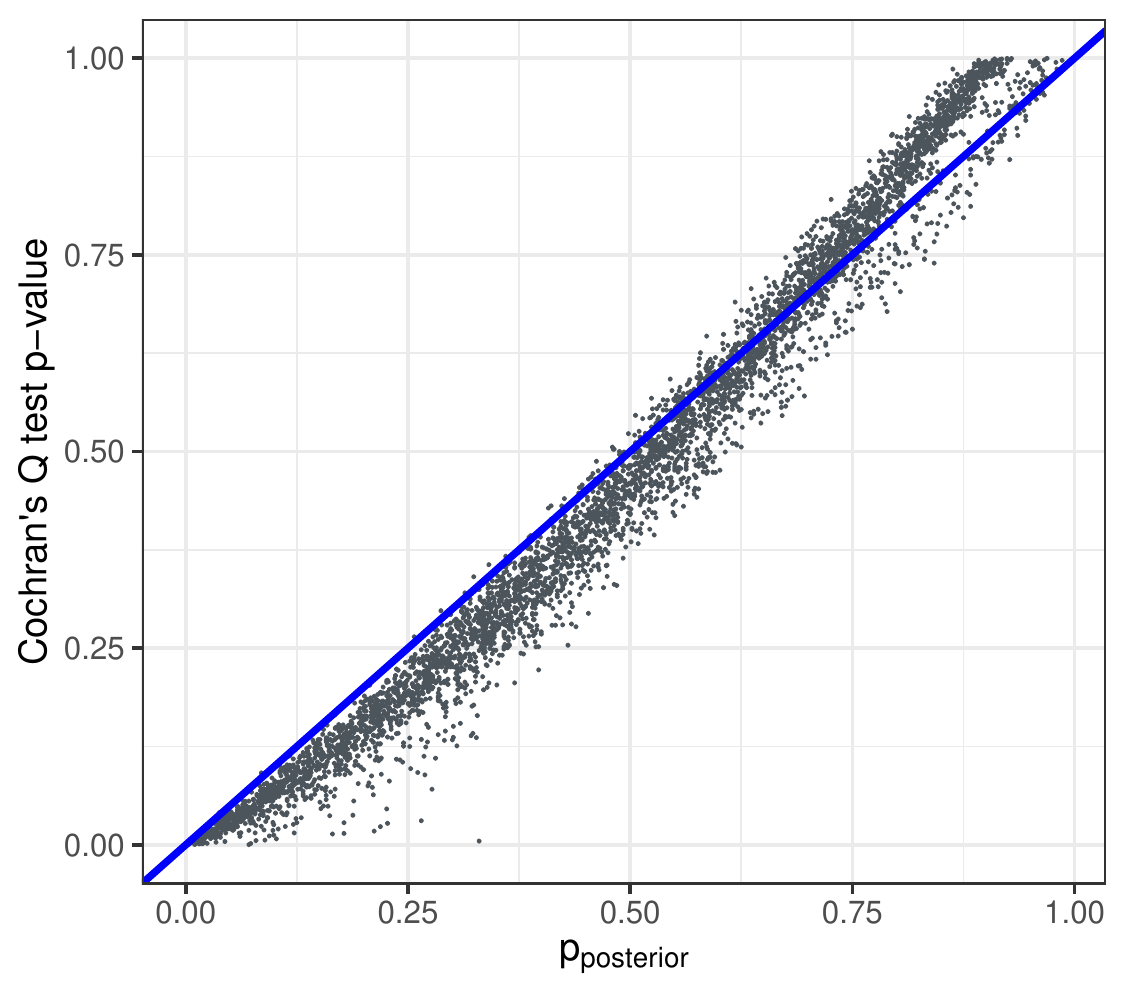}
    \end{subfigure}
    \qquad
      \begin{subfigure}[b]{0.35\textwidth}
       \caption{}
        \includegraphics[width=\textwidth]{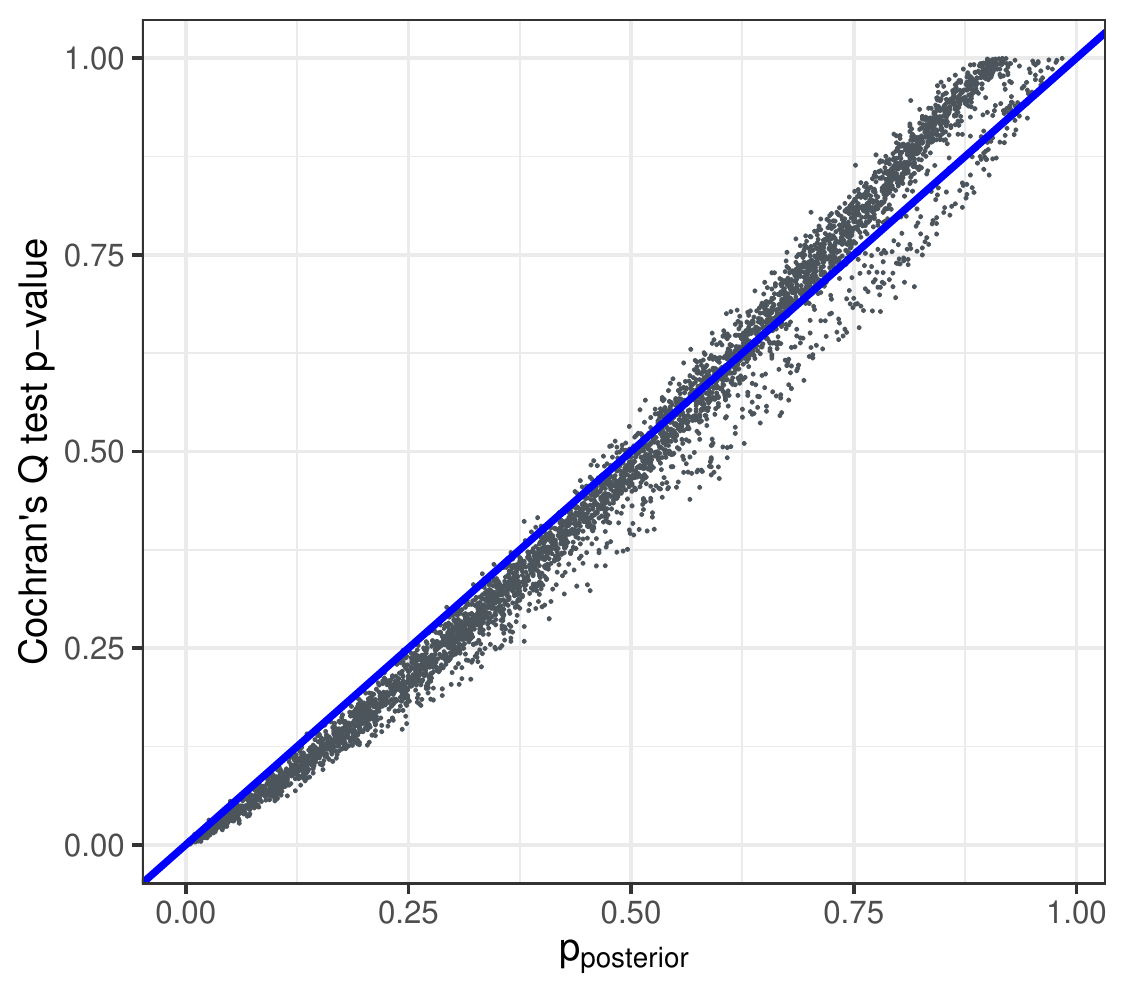}
    \end{subfigure}
         \caption{{\bf Comparison of posterior-PRPs and classic $p$-values for testing heterogeneity under the reference reproducible model~ } Each dataset consists of five experiments without batch effect contamination ($\eta = 0$) but is generated with an intrinsic low level of heterogeneity. Panels {\bf a} and {\bf b} compare the histograms of classic $p$-values, obtained from Cochran's test of heterogeneity, and the proposed posterior-PRPs using the $Q$ quantity. Cochran's $Q$ test assumes a fixed-effect model and is shown sensitive to the low level of heterogeneity.
The posterior-checking procedure, based on the default reference model, tolerates such a level of heterogeneity acceptable. Panels {\bf c} and {\bf d} compare the  two types of $p$-values for each simulated dataset. Overall, they show good agreement. In Panel {\bf c}, the default reference model is used to generate posterior-PRPs. In Panel {\bf d}, the reference model is modified to allow only the fixed-effect,  and the results indicate increased agreement with the classic $p$-values, especially in the range of small values. }\label{comp_Q_prp}
    \end{figure}

We also take the opportunity to empirically compare the Bayesian $p$-values with its frequentist counterpart, the $p$-values derived from Cochran's $Q$ statistic. Our focus is on their behaviors when $\eta = 0$, which relates to specificity. 
Cochran's $Q$ statistics are computed assuming a fixed-effect meta-analysis model, under which they follow a pivotal $\chi^2$ distribution with $m-1$ degrees of freedom. 
Figure \ref{comp_Q_prp} shows that the Bayesian and the frequentist $p$-values are largely aligned. However, the frequentist $p$-values are over-sensitive to the intrinsic heterogeneity within the data, likely because of its stringent fixed-effect assumption.  
To further confirm this point, we re-compute the posterior-PRPs by modifying the reference model to include only the fixed effect assumption (i.e., $\gamma = 0$). 
The resulting Bayesian $p$-values become much more concordant to the frequentist $p$-values, especially for extremely small values (e.g., $p < 0.05$).	   

\subsection{Simulations of Publication Bias}

Our general simulation scheme considers a case-control study of a binary outcome. To introduce selection bias, we impose censoring mechanisms into otherwise normal data generative procedures. 
In all cases, following \cite{Macaskill2001},  the censoring schemes utilize the $p$-values derived from NHSTs.

\subsubsection{Publication Bias Detection in Two-group Scenario}

In the two-group scenario, we simulate two experiments per dataset. Both the original and replication experiments use a balanced case-control design with 200 samples. The true association effects are fixed at odds ratio, $2/3$, in both experiments. The outcomes are generated by a binomial sampling procedure, and the summary statistics (in the form of estimated log-odds ratios and their corresponding standard errors) are obtained by fitting standard logistic regression models.
The censoring scheme only applies to the original experiment. Specifically, a simulated original experiment data is retained only if its association testing $p$-value is smaller than the pre-defined threshold, $p_t$. 
We experiment with $p_t = 0.01, 0.05,$ and 1 (i.e., no censoring). For each threshold value, 5,000 datasets are generated. 

\begin{figure}[H]
\centering
    \begin{subfigure}[b]{0.75\textwidth}
        \includegraphics[width=\textwidth]{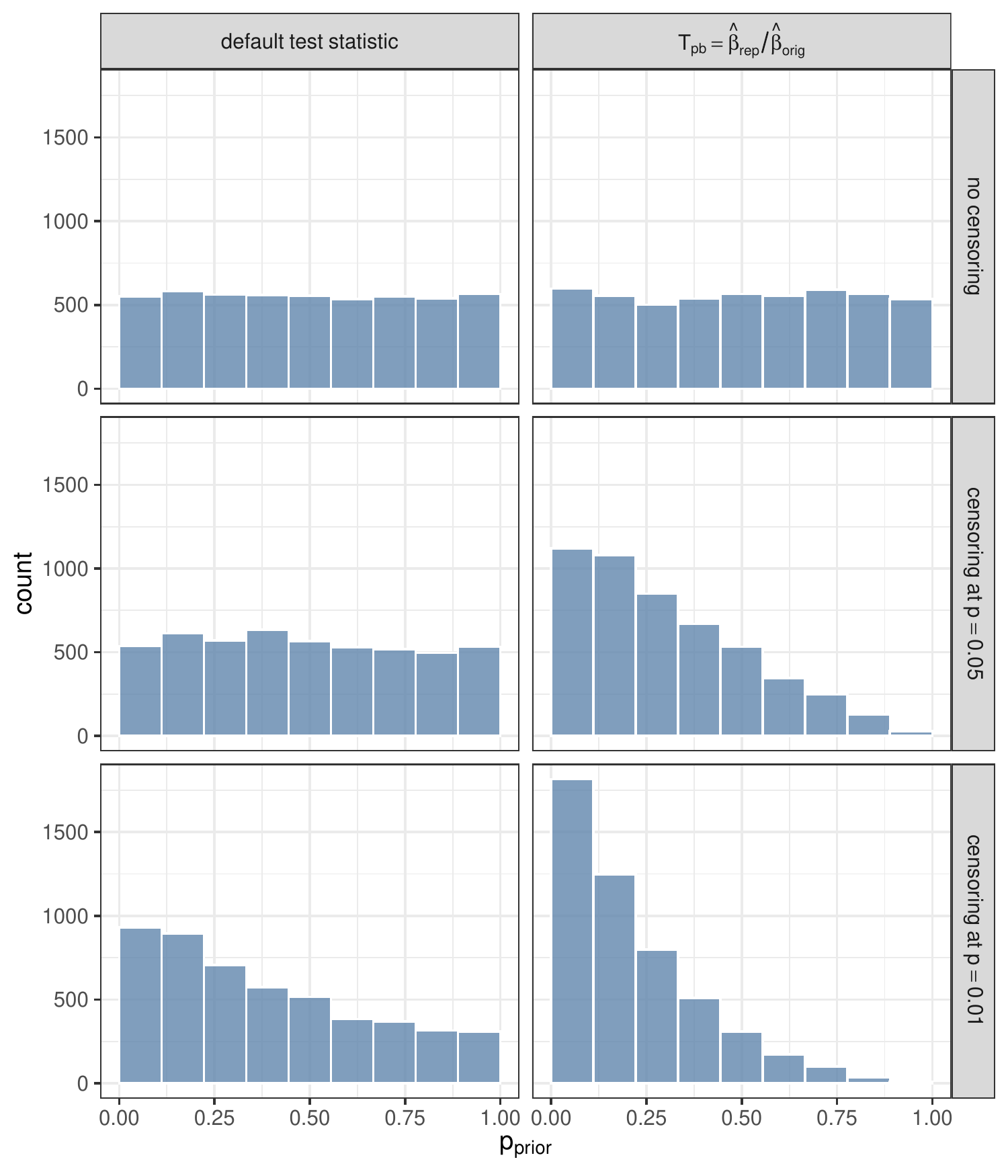}
    \end{subfigure}
    \caption{{\bf Publication bias detection in two-group scenario~} The original experiment data are simulated via a censoring mechanism mimicking publication bias. The replication experiment data are generated without censoring. The left column represents the histograms of the default two-sided prior-PRPs, and the right column represents histograms of the one-sided prior-PRPs designed to detect publication bias. The different rows represent different censoring strengths. Both types of prior-PRPs are well-behaved when publication bias is absent. However, the one-sided $p$-values are clearly more sensitive in detecting publication bias.
}\label{pub2gp}
  \end{figure}

For each simulated dataset, we compute both the default two-sided prior-PRPs and the one-sided values specifically designed for detecting publication bias (i.e., using $T_{\rm pb}$). 
The results are summarized in Figure \ref{pub2gp}.
Both types of $p$-values are seemingly calibrated when there is no selection bias (i.e., $p_t = 1$), and both show positively skewed empirical distributions when the censoring mechanism is in place. 
Nevertheless, the special-purpose replication $p$-values exhibit superior sensitivity over their general-purpose counterpart in the presence of publication bias.

\subsubsection{Publication Bias Detection in Exchangeable-group Scenario}

For the exchangeable-group scenario, our simulations mimic practical settings commonly encountered in meta-analysis and systematic review. 
We consider a dataset consisting of 10 experiments with various sample sizes (5 with 200 samples, 3 with 500 samples, and 2 with 1,000 samples) and balanced case-control designs. 
The true association effects are centered at the odds ratio $= 2/3$ with a low level of heterogeneity (the sign consistency probability = 0.96).
Following \cite{Macaskill2001}, we apply a ``soft" censoring scheme to each participating experiment. Specifically, the simulated data for an experiment is retained with probability, $\exp(-c p^{3/2})$, where $p$ represents the association testing $p$-value and parameter $c$ characterizes the censoring strength. 
We examine strong, modest, and no censoring effects by setting $c = 10, 5$, and 0, respectively. For each $c$ value, 5,000 datasets are generated. 
Given the unequal sample sizes in the 10 participating experiments, this scheme preferentially censors data from small studies and introduces selection bias when $c \ne 0$.   

\begin{figure}[H]
\centering
    \begin{subfigure}[b]{0.85\textwidth}
        \includegraphics[width=\textwidth]{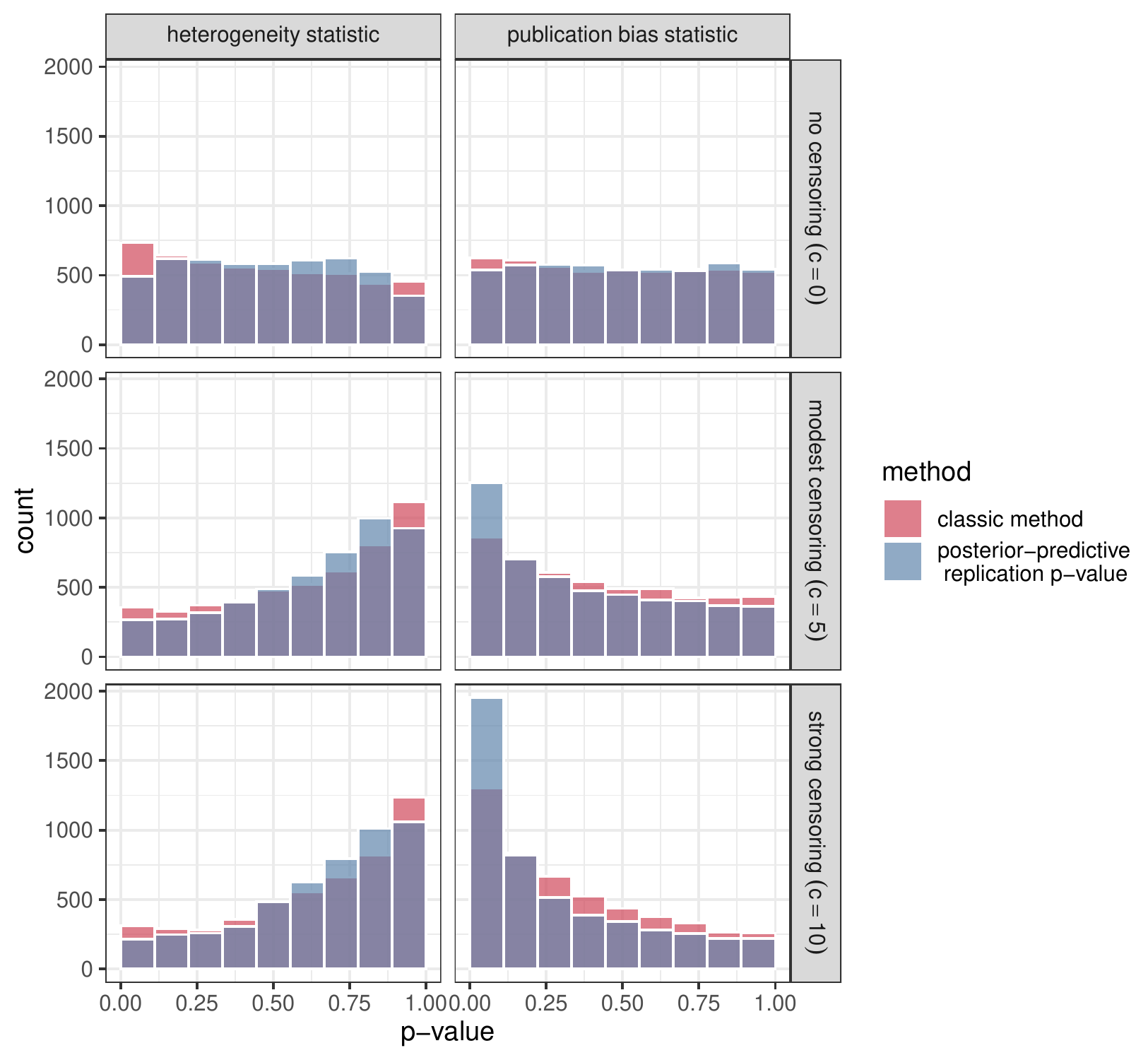}
    \end{subfigure}
    \caption{{\bf Publication bias detection in exchangeable-group scenario~} The right column shows the histograms of posterior-PRPs obtained using Egger's test quantity, overlaid with the histograms of $p$-values from classic Egger regressions.  The left column represents the histograms of posterior-PRPs derived from the $Q$ quantity, overlaid with the histograms of $p$-values from Cochran's $Q$ tests. The strength of the censoring mechanism increases from the top row to the bottom row. While both heterogeneity tests indicate a lack of heterogeneity, the posterior-PRPs designed for detecting publication bias show good sensitivity compared to the classic approach.
}\label{pubmultigp}
  \end{figure}

We analyze each simulated dataset by computing the posterior-predictive replication $p$-values derived from both the $Q$ quantity and Egger regression statistic. 
The results are summarized in Figure \ref{pubmultigp}. 
When selection bias is absent ($c = 0$), there are no excessive small Bayesian $p$-values for either type of test statistic. 
The average $p$-values from both test statistics over 5,000 datasets are close to $1/2$, as expected by the theory.   
In the presence of publication bias, the posterior-predictive replication $p$-values based on Egger test quantity show increased sensitivity as the censoring strength ($c$) increases. 
Interestingly, the $p$-value distributions based on the $Q$ quantity indicate reduced heterogeneity (comparing to the reference model specification), and the $p$-values derived from Cochran's Q test show similar patterns. 
This observation demonstrates that a single diagnosis may not address all issues in replicability. Hence, the assessment of replicability should be multifaceted.
Comparing to the existing frequentist model criticism approaches in detecting funnel plot asymmetry, the proposed approach (with Egger statistic) shows slightly better (but overall similar) sensitivity as the classic Egger regression test.
In this particular simulation setting, it also outperforms the meta-analytic regression test implemented in the R package \texttt{metafor}, which similarly tests the linear trend between $\hat \beta$ and $\sqrt{\sigma^2 + \phi^2}$ (by plugging in a point estimate of $\phi^2$).

\section{Real Data Applications}

\subsection{Re-analysis of RP:P Data}

We apply the prior-predictive $p$-values to re-analyze the RP:P data, whose design falls in the category of the two-group scenario. 
The RP:P project attempts to replicate 100 psychology studies published in three top psychology journals during 2008. 
The replication experiment of each study is designed to match the design, sample size, and analysis methods in the corresponding original experiment. 
Following \cite{Patil2016,  Pawel2020, Hung2020}, we focus our re-analysis on a subset of 73 studies, for which the corresponding effect sizes and standard errors are computed via Fisher's $z$-transformation.

\begin{figure}[H]
\centering
    \begin{subfigure}[b]{0.70\textwidth}
        \includegraphics[width=\textwidth]{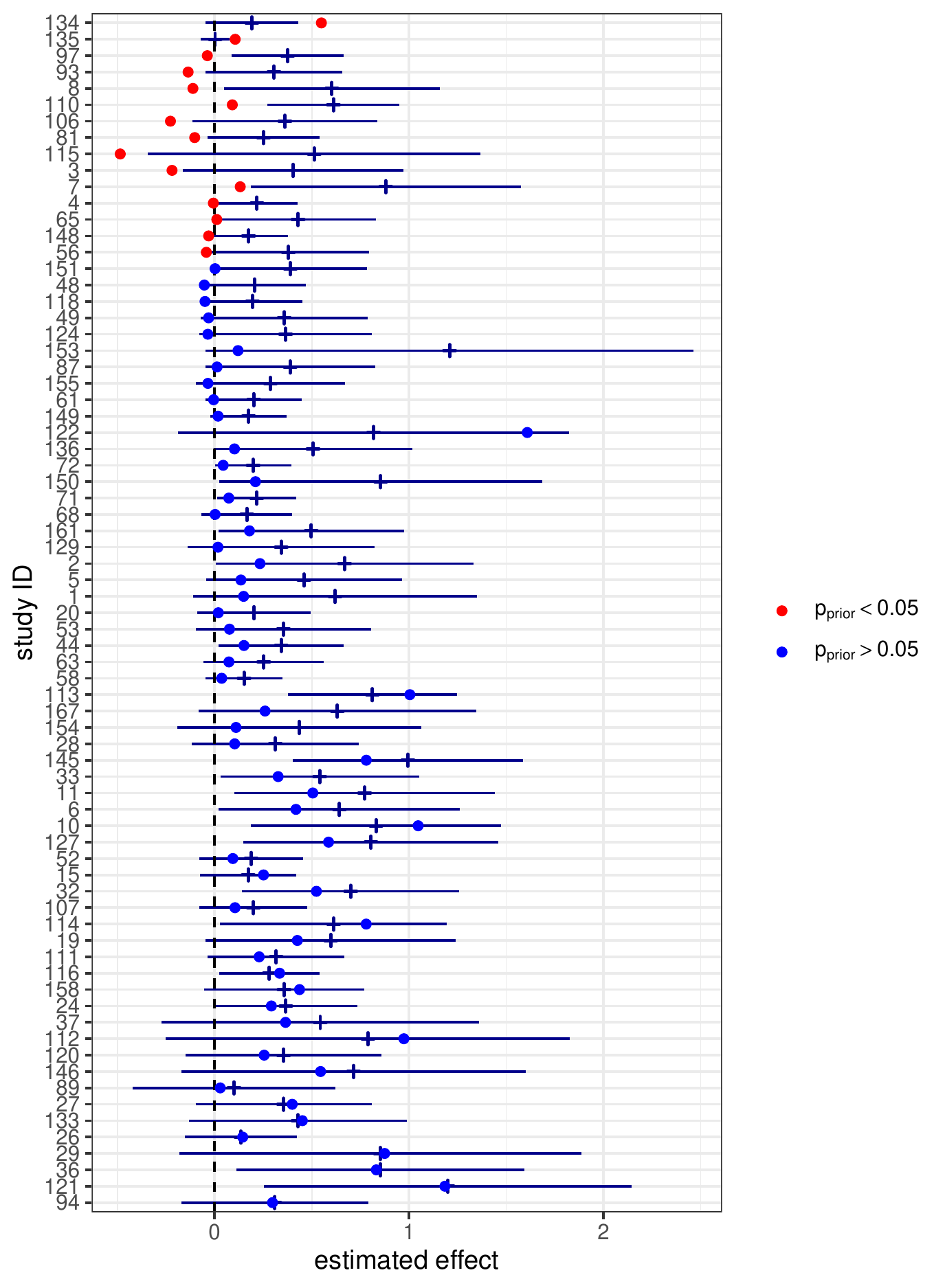}
    \end{subfigure}
    \caption{{\bf Analysis of RP:P data by the prior-checking procedure~} The plot summarizes the analysis result of all 73 studies in the RP:P dataset. The original estimate of effect is represented by a vertical tick centered at the corresponding 95\% predictive interval (horizontal line) for each study. The filled dots are estimates from the replication studies. The red dots are from the studies with the default two-sided prior-PRP $< 0.05$ and do not fall in the corresponding intervals. The studies are sorted by the increasing prior-PRP values. For most participating studies showing small prior-PRPs, the estimated effects tend to shrink toward 0, compared to the original estimates. }\label{rpp.interval.plot}
  \end{figure}

We apply the default reference model and compute the two-sided prior-predictive $p$-values for all 73 studies. 
The result is summarized in Figure \ref{rpp.interval.plot}, where we highlight the consistency between the prior-predictive $p$-values and the Bayesian predictive intervals indicated by Proposition 2. 
15 out of 73 studies are flagged at the 5\% significance level. In comparison, Patil {\it et al.} identify 22 studies by constructing 95\% predictive intervals assuming a fixed-effect meta-analysis model.  
We are able to reproduce their result by resetting our default mixture reference model to a more restrictive special case  (i.e., $\omega^2 \in (10, \infty)\,$ and  $\gamma = 0$).
  
Figure \ref{rpp.interval.plot} also reveals a striking pattern in the RP:P data: the effect estimates in the replications are predominantly shrunk towards 0 in flagged studies. 
This observation prompts computing the one-sided prior-predictive $p$-values based on (\ref{prior_prp.pb}) for detecting publication bias.
This new analysis flags 22 studies at the 5\% significance level, with 13 overlapping with those previously flagged by the two-sided prior-PRPs.
This result suggests that publication bias can be responsible for a large proportion of irreproducible results in psychology research, a conclusion similarly drawn by \cite{Hung2020}.
\begin{figure}[H]
\centering
    \begin{subfigure}[b]{0.7\textwidth}
        \includegraphics[width=\textwidth]{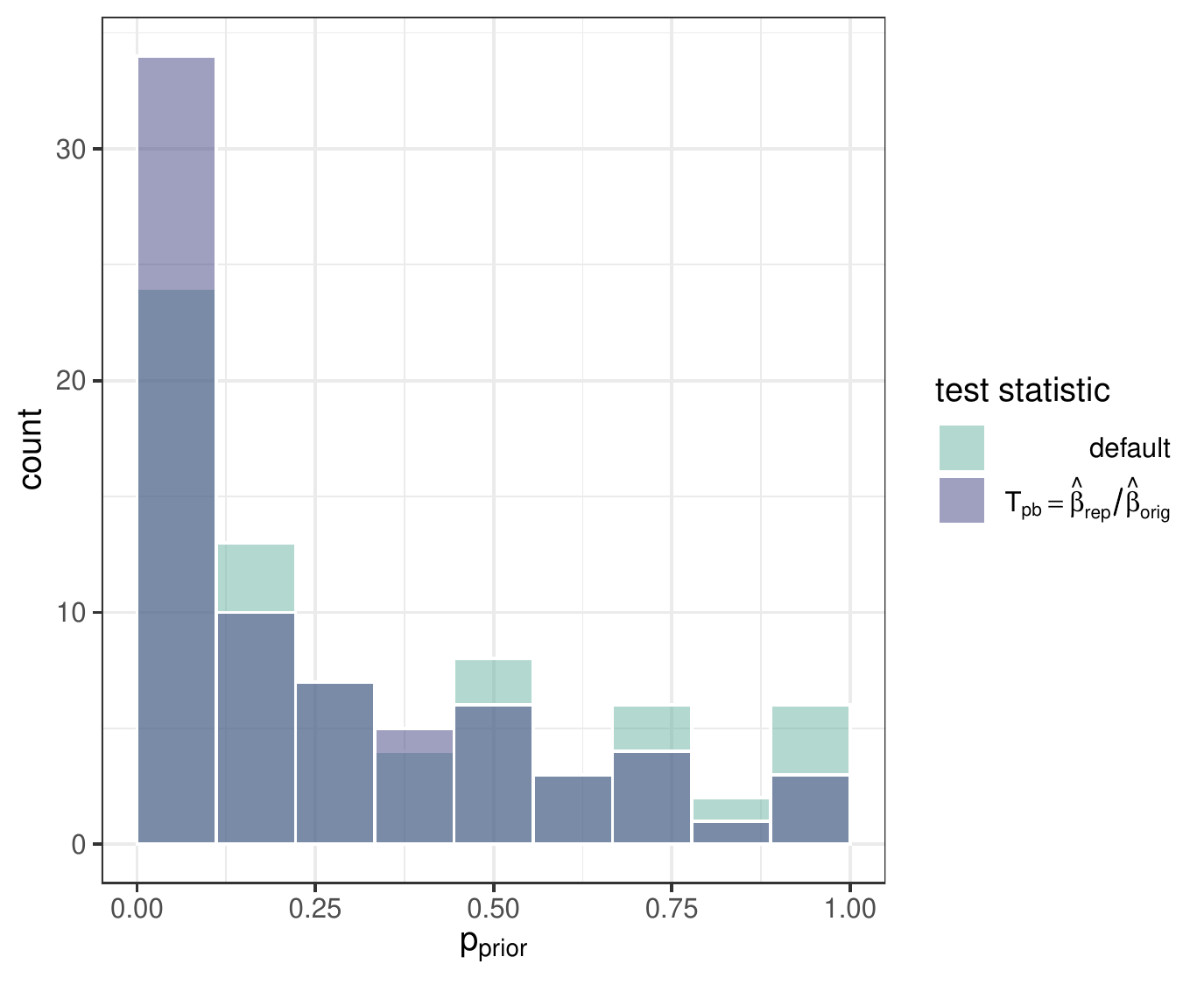}
    \end{subfigure}
    \caption{{\bf Histograms of two types of prior-PRPs in assessing replicability of the RP:P data~} The comparison is between general-purpose and default two-sided prior-PRPs (teal) and the one-sided prior-PRPs specifically designed for detecting publication bias (purple). The two histograms are overlayed, and both show clear positive-skewness. The prior-PRPs based on $T_{\rm pb}$ show more excessive small values and increased skewness.   }\label{rpp.hist}
 \end{figure}
  
Figure \ref{rpp.hist} shows the histograms of two types of prior-predictive replication $p$-values in analyzing RP:P data.
Both empirical distributions show {\em severe} positive-skewness, indicating a worrisome trend in psychology research.   
Such a finding of a global pattern highlights the unique value of the systematic efforts in simultaneously investigating multiple studies by the design of RP:P.       

\subsection{Systematic Review of Impact of Cardiovascular Diseases on Mortality and Severity of COVID-19}

\begin{figure}[H]
\centering
    \begin{subfigure}[b]{0.35\textwidth}
    \caption{Mortality}
        \includegraphics[width=\textwidth]{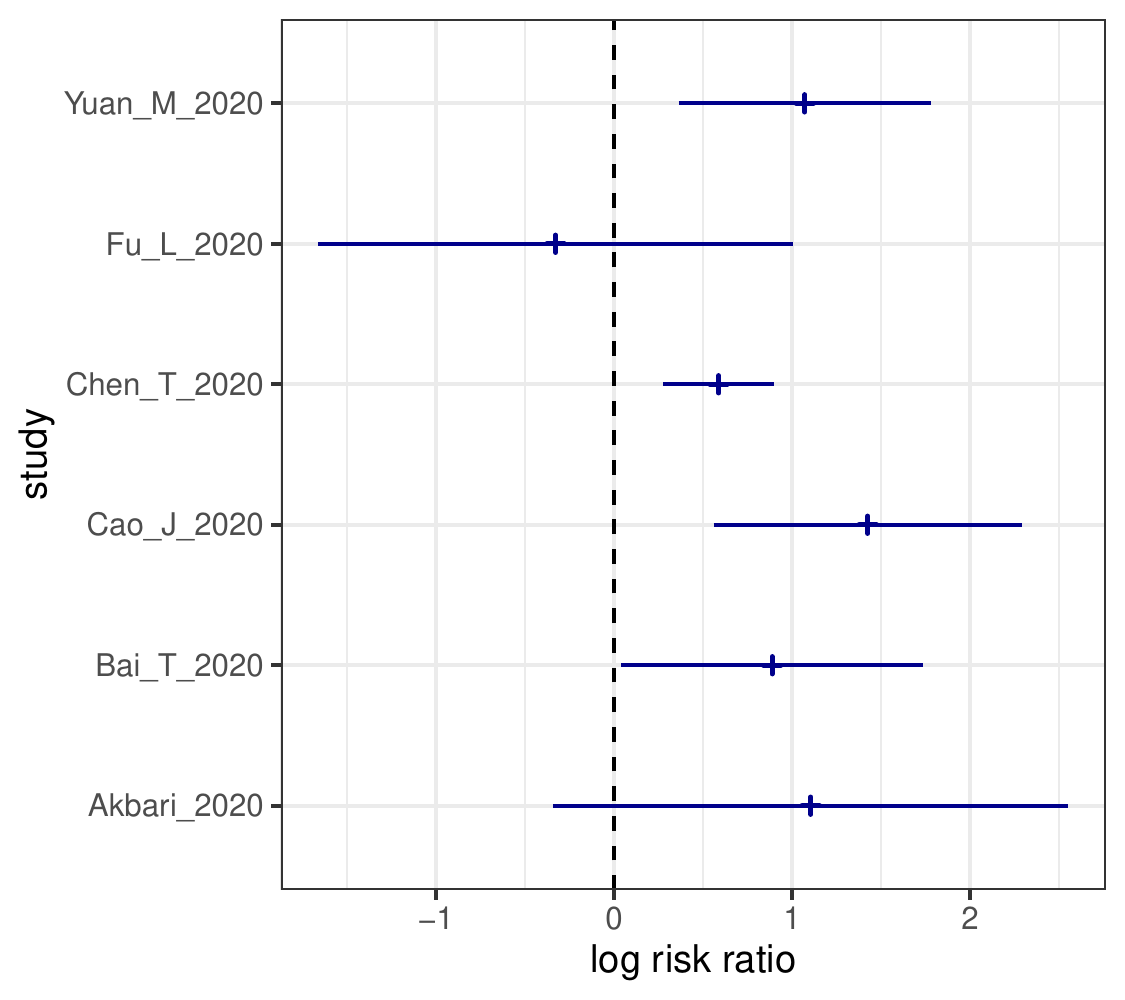}
    \end{subfigure}
    \qquad
      \begin{subfigure}[b]{0.35\textwidth}
       \caption{Severity}
        \includegraphics[width=\textwidth]{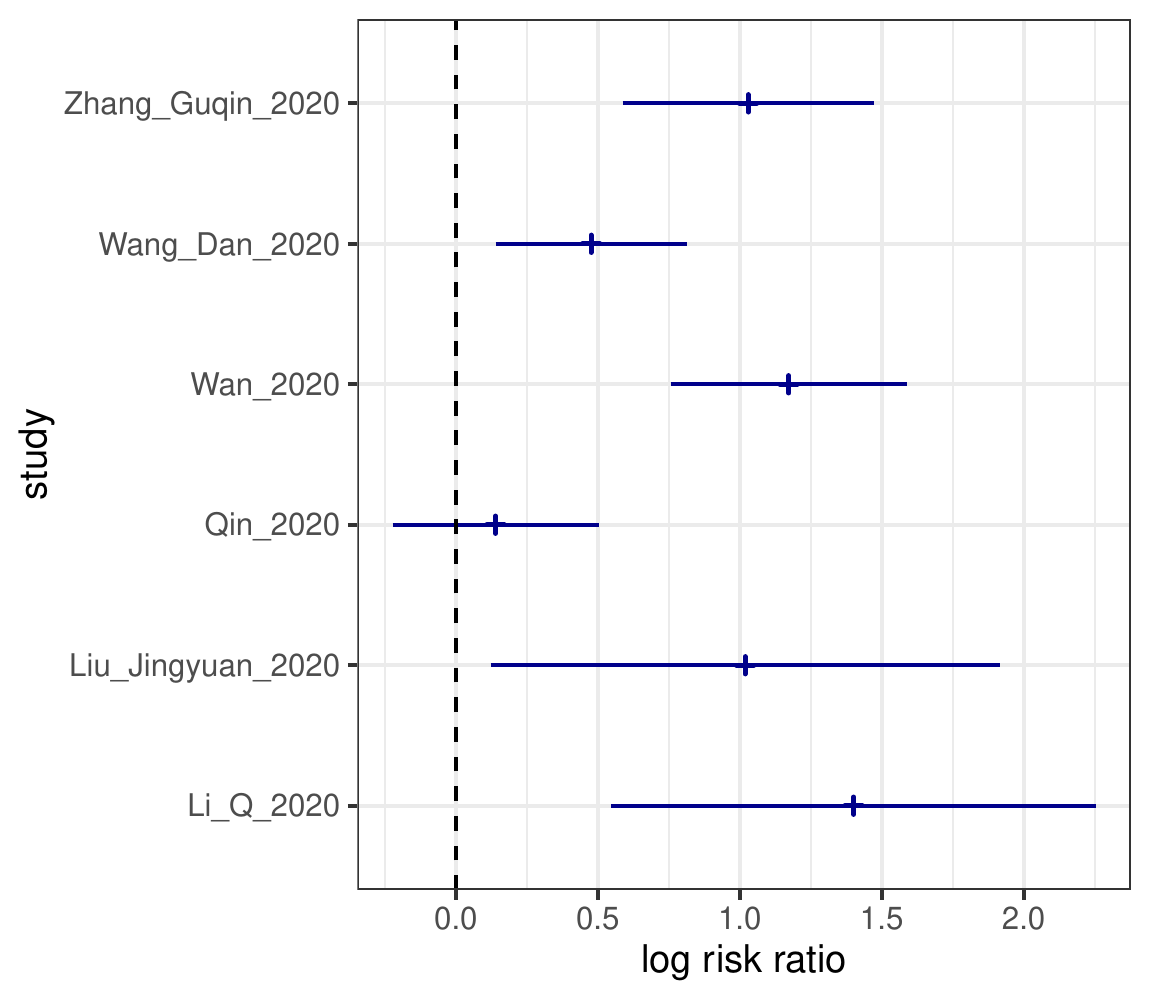}
    \end{subfigure}
    \\
      \begin{subfigure}[b]{0.35\textwidth}
         \caption{Mortality}
        \includegraphics[width=\textwidth]{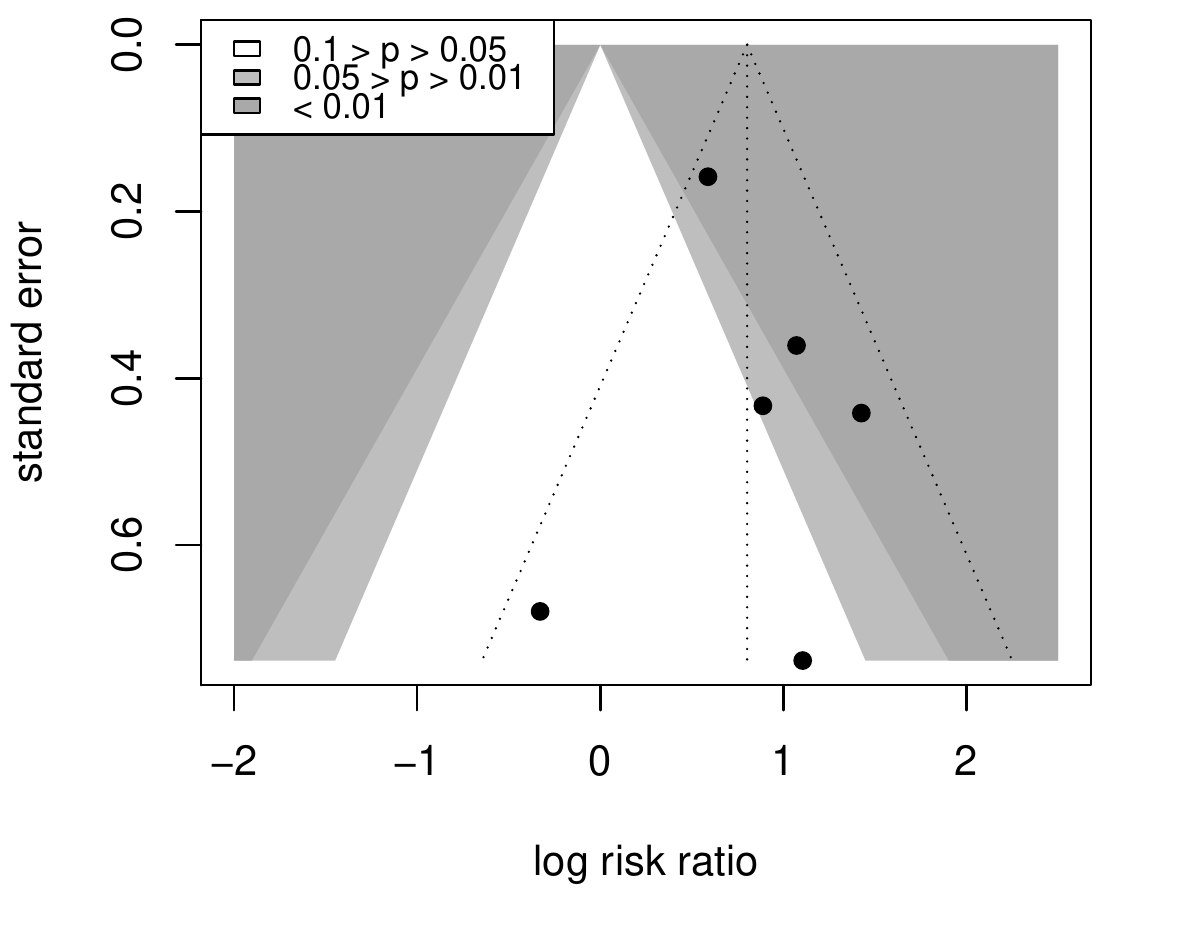}
    \end{subfigure}
    \qquad
       \begin{subfigure}[b]{0.35\textwidth}
         \caption{Severity}
        \includegraphics[width=\textwidth]{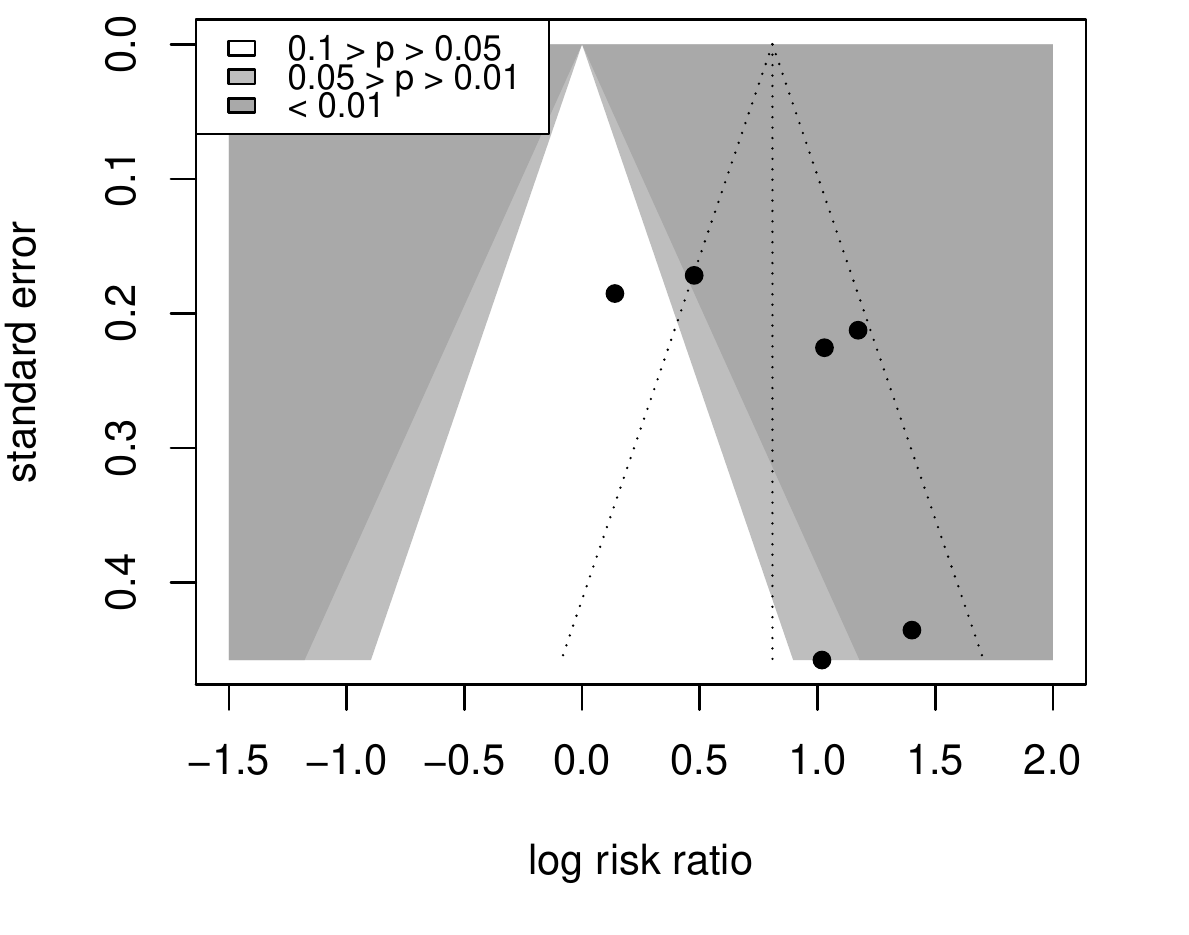}
    \end{subfigure}
    \caption{ {\bf Forest and funnel plots of the COVID-19 datasets ~} Panels {\bf a} and {\bf b} show the forest plots for the estimated effects in the six different experiments on the cardiovascular disease impact the COVID-19 mortality, and the COVID-19 severe case rate, respectively. Panel {\bf c} and {\bf d} show their contour-enhanced funnel plots. }\label{covid}
\end{figure}

Coronavirus disease 2019 (COVID-19), caused by the infection of the SARS-Cov-2 virus, has been linked to a worse prognosis in patients with pre-existing cardiovascular diseases.   
Pranta {\it et al.} \cite{Pranata2020} recently perform a systematic review on association evidence between pre-existing cardiovascular conditions and COVID-19 outcomes.
Six studies investigating the mortality and another six studies investigating the severity of COVID-19 are included in the review.
All 12 studies are observational and retrospective. 
They also have diverse sample sizes, ranging from 24 to 441 (median $=156$).    
Estimates of log-risk ratios and the corresponding standard errors are extracted from each study and are shown in Figure \ref{covid}.

We compute the posterior-predictive replication $p$-values using $Q$ and Egger statistics for both datasets and compare the results to the available traditional analyses described in \cite{Pranata2020}.
The results are summarized in Table \ref{covid.res}. 
\begin{table}[ht]
\centering
{\renewcommand{\arraystretch}{1.25}
\begin{tabular}{c c c c c c c}
\hline
 \multirow{2}*{COVID-19 Outcome} & ~ & \multicolumn{2}{c}{Test of Heterogeneity}& ~ &\multicolumn{2}{c}{Test of Publication Bias}  \\
\cline{3-4} \cline{6-7}
  & & classic & posterior-PRP & ~~~ & classic & posterior-PRP \\
  \hline \hline
Mortality & & $0.221$ & $0.278$ &~~ & $0.429$ & $0.598$ \\
 \hline 
Severity & &  $ 0.001$ & $ 0.010 ~~$ & & $0.010$ & $0.240$ \\

\hline 
\end{tabular}}
\vskip 3pt
\caption{{ \bf Systematic review of pre-existing cardiovascular disease on COVID-19 outcomes } The table shows the comparisons between the classic approaches and the proposed posterior-predictive replication $p$-values in assessing excessive heterogeneity and potential publication bias. The classic $p$-values are obtained from Cochran's $Q$ test for testing heterogeneity, and the posterior-PRPs are computed using the $Q$ quantity. For detecting publication bias, the classic $p$-values are obtained from Egger regression test implemented in R package \texttt{metafor}, and the posterior-PRPs are computed using Egger statistic with sampled hyperparameters.}\label{covid.res}
\end{table}

For heterogeneity assessment, Cochran's tests and the Bayesian $p$-values are qualitatively similar in both mortality and severity studies. 
As expected, the Bayesian $p$-values are more conservative by allowing reasonable heterogeneity in the reference model.  
In all test cases, only the severity study is flagged by the posterior-predictive replication $p$-value based on $Q$ quantity at 5\% significance level (Bayesian $p$-value $=0.010$, Cochran's $\chi^2$ test $p$-value $= 0.001$).
Upon close inspection, the larger-than-expected heterogeneity is seemingly driven by a single study with a large sample size (Qin 2020) in the analysis, which under-estimates the effect size than the remaining studies. 
(The posterior-PRP based on the remaining five studies is 0.157.) 
The only disagreement between the classic and the proposed assessment approaches appears in detecting publication bias for the severity outcome. The Bayesian PRP based on Egger regression quantity does not deem the data to provide strong enough evidence against reproducibility.  Figure \ref{covid} does not seem to show a clear linear trend between the estimates and the corresponding standard errors upon close examination. 
As a rule of thumb, it is commonly recommended that the classic Egger regression test should be applied to $ \ge 10$ studies, partially because of the calibration of the null distribution and the considerations for lack of sensitivity  \cite{Handbook}.  (Pranata {\it et al.} do not formally list the classic Egger regression result as numerical evidence but briefly comment on potential funnel plot asymmetry.) 
In this particular case, the Bayesian $p$-value solution shows some advantages without requiring a pre-defined null distribution.     
However, the second concern remains, and we caution the interpretation that the publication bias is {\em absent}.
We should conclude that the potential publication bias may not be strong enough to be flagged by the given experiments using the proposed method.

\section{Discussion}

Applying model criticism approaches in assessing replicability is not new. The uses of traditional Cochran's $Q$ test and Egger regression in exchangeable-group scenarios all fall in this statistical framework.  
A common limitation of these existing (frequentist) methods is that the reference (i.e., the null) model is often too restrictive to be realistic (e.g., the fixed-effect assumption). 
We address this limitation by adopting more general Bayesian hierarchical models while following the same model criticism principles. 
The existing toolkit built for Bayesian model criticism has its unique, pragmatic flexibility to adopt the extended reference models and make inferences in the presence of nuisance parameters.
More importantly, it is versatile in dealing with different application scenarios arising from replicability assessment.

A key element in replication assessment is to define replication success. We argue that this definition should be ultimately context-dependent. 
Nevertheless, we believe definitions based on the DC criterion can be useful if prior established quantitative standards for reproducibility lack in some scientific domains. 
In other words, replication definitions based on the DC criterion can serve as a reasonable starting point, and other domain-specific standards should gradually replace it with the accumulation of new data and knowledge. 
Additionally, our presented model is not the only way to set up the reference model for replicability assessment. Our choice to present this particular version of the reference model is due to its connections to the commonly-used random-effect meta-analysis model.
Thus, it is more convenient to compare the proposed Bayesian model criticism approaches and the traditional approaches assuming similar models. 
The curved exponential family normal (CEFN) model \cite{Wen2014, Zhao2020} is another parametric model capable of faithfully implementing the DC criterion.

A well-known drawback of model criticism approaches is that their conclusions are often misinterpreted and misused. 
The binary outcome from a model criticism procedure is either flagging/rejecting a reference model or not. 
If the corresponding reference model is flagged, the correct interpretation is via Fisher's disjunction. 
A common mistake is to misinterpret a lack of evidence for flagging as evidence for supporting the reference model, and such a mistake is not unique for replication assessment. 
This is why we wish to highlight the importance of the non-informativeness principle and its implications in our proposed procedures: it serves as a good reminder to avoid misinterpreting the proposed replication $p$-values in assessing replicability.

Model criticism is not the only statistical inference framework capable of assessing replicability. For example, inference procedures based on model comparison principles are also applicable. 
A model comparison approach aims to classify the observed data into mutually exclusive latent data-generative models, e.g., irreproducible vs. reproducible models.
%The outcome of a model selection procedure is also straightforward to interpret. 
Our previous work \cite{Zhao2020} and the influential work by Li {\it et al.} \cite{Li2011} are both  Bayesian model comparison procedures for replicability assessment in high-throughput experiments.    
Nevertheless, we note that prior quantification on data-generative models is critical for the success of model comparison procedures.
In high-throughput settings, this information can be sufficiently ``learned" from data by partial pooling, whereas in other classic settings of replicability assessment, justifying particular prior choices can be challenging. 
In comparison, the model criticism strategy presents some attractive and practical simplicity: it does not require a prior specification on generative models, nor does it requires a parametric specification of an irreproducible model. 
For many applications discussed in this paper, this operational simplicity is highly desired.
In summary, we believe both model criticism and model comparison approaches can be useful for different application settings; they can be complementary in some cases. 
It is also important to realize that, while they represent different inference strategies, they share the same guiding principles for replication assessment (or even the underlying probabilistic generative models).

\clearpage

\bibliographystyle{naturemag}

\clearpage

\begin{appendices}

\section{Construction of Reference Reproducible Model} \label{prior.spec}

The main text has detailed the specification of set $\Gamma$, which defines various levels of tolerable heterogeneity for the reference model. 
Here we describe the construction of set $\Omega$, which specifies the prior effect size of $\bbar$.
Intentionally, we set up the grid values of $\omega$ to be compatible with the observed data, such that the potential poor fitting of the reference model can only be attributed to the incompatibility of the heterogeneity parameter, $\gamma$. 

This general principle is applied to both prior and posterior checking procedures, but the implementations are slightly different.
In both cases, instead of directly setting $\Omega$, we first specify a grid of values for $\lambda^2 = (\phi^2 + \omega^2)$, then compute $\omega^2 = (1- \gamma) \, \lambda^2$. 

\subsection*{Two-group Scenario}
 
In this scenario, we determine $\Lambda$, the set of $\lambda^2$ values, by $\hbo$ and $\sigma_{\rm orig}$. Let $q_1, q_2, q_3$ denote the first, second, and the third quartiles from the $\chi^2_1$ distribution. 
We regard ${\hat \beta^2}_{\rm orig}+\sigma_{\rm orig}^2$ as a proxy for an estimate of $\lambda^2$ and reason that $\frac{{\hat \beta^2}_{\rm orig}+\sigma_{\rm orig}^2}{\lambda^2}$ should be a {\em plausible} draw from the $\chi^2_1$ distribution. 
Thus, we construct 
\begin{equation*}
  \Lambda = \left \{ \frac{{\hat \beta^2}_{\rm orig}+\sigma_{\rm orig}^2}{q_3},\, \frac{{\hat \beta^2}_{\rm orig}+\sigma_{\rm orig}^2}{q_2}, \, \frac{{\hat \beta^2}_{\rm orig}+\sigma_{\rm orig}^2}{q_1} \right \}
\end{equation*}

\subsection*{Exchangeable-group Scenario}

In this scenario, we first compute a fixed-effect estimate of $\bbar$ , i.e.,
\begin{equation*}
  \hat \bbar = \frac{ \sum_{i=1}^m w_i \hat \beta_i}{\sum_{i=1}^m  w_i},~\mbox { where } w_i = \frac{1}{\sigma_i^2},
\end{equation*}
and
\begin{equation*}
  {\rm se}(\hat \bbar) = \sqrt{\frac{1}{\sum_{i=1}^m  w_i}}.
\end{equation*}
We construct set $\Lambda$ by
\begin{equation*}
  \Lambda = \left \{ \left( \hat \bbar - {\rm se}(\hat \bbar)\right)^2, ~ {\hat \bbar}^2, ~ \left( \hat \bbar + {\rm se}(\hat \bbar) \right)^2 \right \}.
\end{equation*}

\section{Proof of Proposition 1}\label{prop1.proof}
\begin{proof}
We introduce a shorthand notation, $P \circ T(\Xv_{\rm orig}, \Xv_{\rm rep})$, to represent prior-predictive replicaiton $p$-value, $\Pr(T(\Xv)\leq T(\Xv_{\rm rep})|\Xv_{\rm orig},\Xv_{\rm rep},\Mr)$ and emphasize it as a function of random variables $\Xv_{\rm orig}$ and $\Xv_{\rm rep}$. 

It follows that
\begin{equation*}
 \begin{aligned}
  & ~~~ \Pr\left[\, P \circ T (\Xv_{\rm orig}, \Xv_{\rm rep}) \le p \, \mid \, \Mr \, \right ] \\
  & =   \Pr\left[\, \Pr \left ( P \circ T (\Xv_{\rm orig}, \Xv_{\rm rep}) \le p \, \mid \Xv_{\rm orig}, \Mr \right) \, \mid \, \Mr \, \right ]
 \end{aligned}
\end{equation*}
Let $F_{T\, | \Xv_{\rm orig}, \Mr}$ denote the continuous cumulative distribution function of the predictive distribution, $T(\Xv) \mid \Xv_{\rm orig}, \Mr$. Note that,
\begin{equation*}
\begin{aligned}
  & ~~~ \Pr \left ( P \circ T (\Xv_{\rm orig}, \Xv_{\rm rep}) \le p \, \mid \Xv_{\rm orig}, \Mr \right) \\
  & = \Pr \left( T(\Xv_{\rm rep}) \le {F^{-1}}_{T\, | \Xv_{\rm orig}, \Mr} (p) \, \mid \, \Xv_{\rm orig}, \Mr \right) \\
  & = p
\end{aligned}
\end{equation*}
Thus,
\begin{equation*}
\Pr\left[\, P \circ T (\Xv_{\rm orig}, \Xv_{\rm rep}) \le p \, \mid \, \Mr \, \right ]  = p.
\end{equation*}
That is, the prior-predictive replication $p$-values are uniformly distributed under the resampling of $\Xv_{\rm orig}$ and $\Xv_{\rm rep}$.

\end{proof}

\section{Proof of Proposition 2}\label{prop2.proof}

\begin{proof}

\begin{equation*}
  \begin{aligned}
& ~~~~ T(\Xv_{\rm rep}) \not\in \left ( q_{\alpha/2}(T(\Xv)\,\mid \, \Xv_{\rm orig}, \Mr )\,,\, q_{(1-\alpha/2)}(T(\Xv)\,\mid \, \Xv_{\rm orig}, \Mr )  \right) \\
 & \iff  T(\Xv_{\rm rep}) \le q_{\alpha/2}(T(\Xv) ~~~ \mbox{or}~~~  T(\Xv_{\rm rep}) \ge q_{1-\alpha/2}(T(\Xv)) \\
 & \iff  \min \Big \{ \Pr( \,T(\Xv) \ge T(\Xv_{\rm rep}) \mid \Xv_{\rm orig}, \Xv_{\rm rep}, {\rm M_R} \,)\,, \\
 &~~~~~~~~~~~~~\Pr( \,T(\Xv) \le T(\Xv_{\rm rep}) \mid \Xv_{\rm orig}, \Xv_{\rm rep}, {\rm M_R} \,)\, \Big \} \le \alpha/2 \\
  & \iff  p_{\rm prior} \le \alpha 
  \end{aligned} 
\end{equation*}

\end{proof}
\section{Computation of Prior-predictive Replication $p$-values}\label{prior-prp.derive}

As described in the main text, we consider $K$ different $(\omega_k, \gamma_k)$ hyperparameter values with equal prior probabilities.
Correspondingly, $\phi_k^2 =  \omega_k^2 \, \gamma/(1-\gamma)$.  

Given $(\omega_k, \phi_k)$, the posterior distribution of $\bbar$ is 
\begin{equation*}
\bbar\,|\, \hbo, \omega_k, \phi_k \, \sim {\rm N}\left(\left(\frac{1}{\sigma_{\rm orig}^2+\phi_k^2}+\frac{1}{\omega_k^2}\right)^{-1}\left(\frac{\hbo}{\sigma_{\rm orig}^2+\phi_k^2}\right) \, , \,\left(\frac{1}{\sigma_{\rm orig}^2+\phi_k^2}+\frac{1}{\omega_k^2}\right)^{-1}\right)
\end{equation*}
The desired posterior predictive distribution is a mixture normal, i.e.,
\begin{equation*}
\hat \beta \,|\, \hbo, \Mr \sim \sum_k w_k \cdot {\rm N} \left(\left(\frac{1}{\sigma_{\rm orig}^2+\phi_k^2}+\frac{1}{\omega_k^2}\right)^{-1}\left(\frac{\hbo}{\sigma_{\rm orig}^2+\phi_k^2}\right)\,,\,\left(\frac{1}{\sigma_{\rm orig}^2+\phi_k^2}+\frac{1}{\omega_k^2}\right)^{-1}+\phi_k^2+\sigma_{\rm rep}^2\right),
\end{equation*}
where
\begin{equation*}ß
w_k = \frac{\mathcal{N} ( \hbo\, \mid \, 0 \,,\, \omega_k^2+\phi_k^2+\sigma_{\rm orig}^2)}{\sum_{j=1}^K { \mathcal{N} ( \hbo\, \mid \, 0 \,,\, \omega_j^2+\phi_j^2+\sigma_{\rm orig}^2)}},
\end{equation*}
and $\mathcal{N} ( \hbo\, \mid \, 0 \,,\, \omega_k^2+\phi_k^2+\sigma_{\rm orig}^2)$ denote a normal density function with mean $0$ and variance $\omega_k^2+\phi_k^2+\sigma_{\rm orig}^2$ evaluated at $\hbo$.

\section{Computation of Posterior-predictive replication $p$-values} \label{posterior-prp.derive}

Here we detail the sampling scheme described in Algorithm 1 for the reference reproducible model.

As described in the main text, we consider $K$ different $(\omega_k, \gamma_k)$ hyperparameter values with equal prior probabilities.
Correspondingly, $\phi_k^2 =  \omega_k^2 \, \gamma/(1-\gamma)$.  
The marginal likelihood of observed data for $(\omega_k, \phi_k)$ is represented by $\mathcal{N}(\hat{\bv} \,\mid\, {\bf 0},\,\Sv_k)$, where $\Sv_k$ is an $m \times m $ covariance matrix and
\begin{equation*}
\Sv_k = 
\begin{bmatrix}
    \omega_k^2 + \phi_k^2 + \sigma_1^2       & \omega_k^2 & \dots & \omega_k^2 \\
    \omega_k^2   &  \omega_k^2 + \phi_k^2 + \sigma_2^2 & \dots & \omega_k^2 \\
    \vdots & \vdots & \ddots & \vdots \\
     \omega_k^2   &  \omega_k^2 & \dots  & \omega_k^2 + \phi_k^2 + \sigma_m^2
\end{bmatrix}
\end{equation*}

The key procedure in the sampling scheme is to draw from the posterior, $P(\tv \mid \hat{\bv}, \Mr)$. Specifically, 
\begin{enumerate}
 \item Draw a single generative model $(\omega_k, \phi_k)$ from the $K$ candidates with  probability
 \begin{equation*}
    p_k = \frac{\mathcal{N}(\hat{\bv} \,\mid\, {\bf 0},\,\Sv_k)}{\sum_{j=1}^K \mathcal{N}(\hat{\bv} \,\mid\, {\bf 0},\,\Sv_j)}
 \end{equation*}

 \item Conditional on $(\omega_k, \phi_k)$ and $\hat \bv$, draw $\bbar$ by
 \begin{equation*}
\bbar\, |\, \hat{\bv}, \omega_k, \phi_k \sim  {\rm N} \left(\left(\frac{1}{\omega_k^2}+\sum_{j=1}^m \frac{1}{\sigma_j^2+\phi_k^2}\right)^{-1}\left(\sum_{j=1}^m\frac{\hat{\beta}_j}{\sigma_j^2+\phi_k^2}\right),\left(\frac{1}{\omega_k^2}+\sum_{j=1}^m \frac{1}{\sigma_j^2+\phi_k^2}\right)^{-1}\right)
\end{equation*}
 \item Draw $\beta_j, j=1,...,m$ by
 \begin{equation*}
\beta_j\,|\,\hat{\bv},  \bbar, \omega_k, \phi_k \sim {\rm N} \left( \left( \frac{1}{\phi_k^2}+\frac{1}{\sigma_j^2}\right)^{-1}\left(\frac{\bar{\beta}}{\phi_k^2}+\frac{\hat{\beta}_j}{\sigma_j^2}\right),\left( \frac{1}{\phi_k^2}+\frac{1}{\sigma_j^2}\right)^{-1}\right)
\end{equation*} 

\end{enumerate}

Finally, given all $\beta_j$'s, we re-sample the observed data by 
\begin{equation*}
\hat{\beta}_j' \, \mid \, \beta_j \sim {\rm N} \left( \beta_j ,\sigma_j^2 \right).
\end{equation*}

\end{appendices}

\end{document}